%% file: _main.tex
\pdfoutput=1

\documentclass[11pt]{article}

\usepackage{ACL2023}

\usepackage{times}
\usepackage{latexsym}
\usepackage{graphicx}
\usepackage{booktabs}
\usepackage{pifont}
\usepackage[T1]{fontenc}
\usepackage[normalem]{ulem}
\usepackage{multirow}
\usepackage{tikz}
\usepackage{soul}
\usepackage{fontawesome5}

\newcommand{\checkmark}{\ding{51}}%

\newcommand{\xhdr}[1]{\vspace{1.5mm}\noindent{{\bf #1.}}}

\usepackage[utf8]{inputenc}
\usepackage{tabularx}

\usepackage{microtype}
\usepackage{xcolor}
\definecolor{lightgray}{HTML}{e6e6e6}

\usepackage{inconsolata}
\newcommand{\ourmodel}{\textsc{Imbue}}
\newcommand{\ourmodels}{\textsc{Imbue }}

\title{\textsc{Imbue}: Improving Interpersonal Effectiveness through Simulation and Just-in-time Feedback with Human-Language Model Interaction}

\author{Inna Wanyin Lin\textsuperscript{1}\ \ \ \ Ashish Sharma\textsuperscript{1} \ \ \ \ Christopher Michael Rytting\textsuperscript{1} \ \ \ \ \\{\bf Adam S. Miner\textsuperscript{2}} \ \ \ \  {\bf Jina Suh\textsuperscript{3}} \ \ \ \ {\bf Tim Althoff\textsuperscript{1}}  \\
\textsuperscript{1}Paul G. Allen School of Computer Science \& Engineering, University of Washington \\
\textsuperscript{2}Stanford University 
\textsuperscript{3}Microsoft Research \\
\texttt{ilin@cs.washington.edu}}

\newcommand{\inna}[1]{\textcolor{magenta}{}}
\newcommand{\tim}[1]{\textcolor{red}{}}
\newcommand{\chris}[1]{\textcolor{blue}{}}
\newcommand{\ashish}[1]{\textcolor{teal}{}}
\newcommand{\jina}[1]{\textcolor{orange}{}}

\begin{document}

\maketitle
\input{00_abstract}

\input{01_introduction}

\input{02_framework}
\input{03_data}

\input{04_method}

\input{05_experiments}

\input{06_user_study}

\input{07_related_work}

\input{08_conclusion}

\input{09_limitations_ethics_statement}

\bibliography{anthology,custom}
\bibliographystyle{acl_natbib}


\input{10_Appendix}
\label{sec:appendix}

\end{document}

%% file: 00_abstract.tex
\begin{abstract}

Navigating certain communication situations can be challenging due to individuals' lack of skills and the interference of strong emotions.
However, effective learning opportunities are rarely accessible.
In this work, we conduct a human-centered study that uses language models to \textit{simulate} bespoke communication training and \textit{provide just-in-time feedback} to support the practice and learning of interpersonal effectiveness skills. 
We apply the interpersonal effectiveness framework from Dialectical Behavioral Therapy (DBT), DEAR MAN, which focuses on both conversational and emotional skills. 
We present \ourmodel, an interactive training system 
that provides feedback 25\% more similar to experts' feedback, compared to that generated by GPT-4.
\ourmodels is the first to focus on communication skills and emotion management simultaneously, incorporate experts' domain knowledge in providing feedback, and be grounded in psychology theory. Through a randomized trial of 86 participants, we find that \ourmodel's simulation-only variant significantly improves participants' self-efficacy (up to 17\%) and reduces negative emotions (up to 25\%). With \ourmodel's additional just-in-time feedback, participants demonstrate 17\% improvement in skill mastery, along with greater enhancements in self-efficacy (27\% more) and reduction of negative emotions (16\% more) compared to simulation-only. The improvement in skill mastery is the only measure that is transferred to new and more difficult situations; situation specific training is necessary for improving self-efficacy and emotion reduction.

\end{abstract}

%% file: 01_introduction.tex
\section{Introduction}
\begin{figure*}[t!]
    \centering
    \vspace{-3mm}
    \includegraphics[width=1\linewidth]{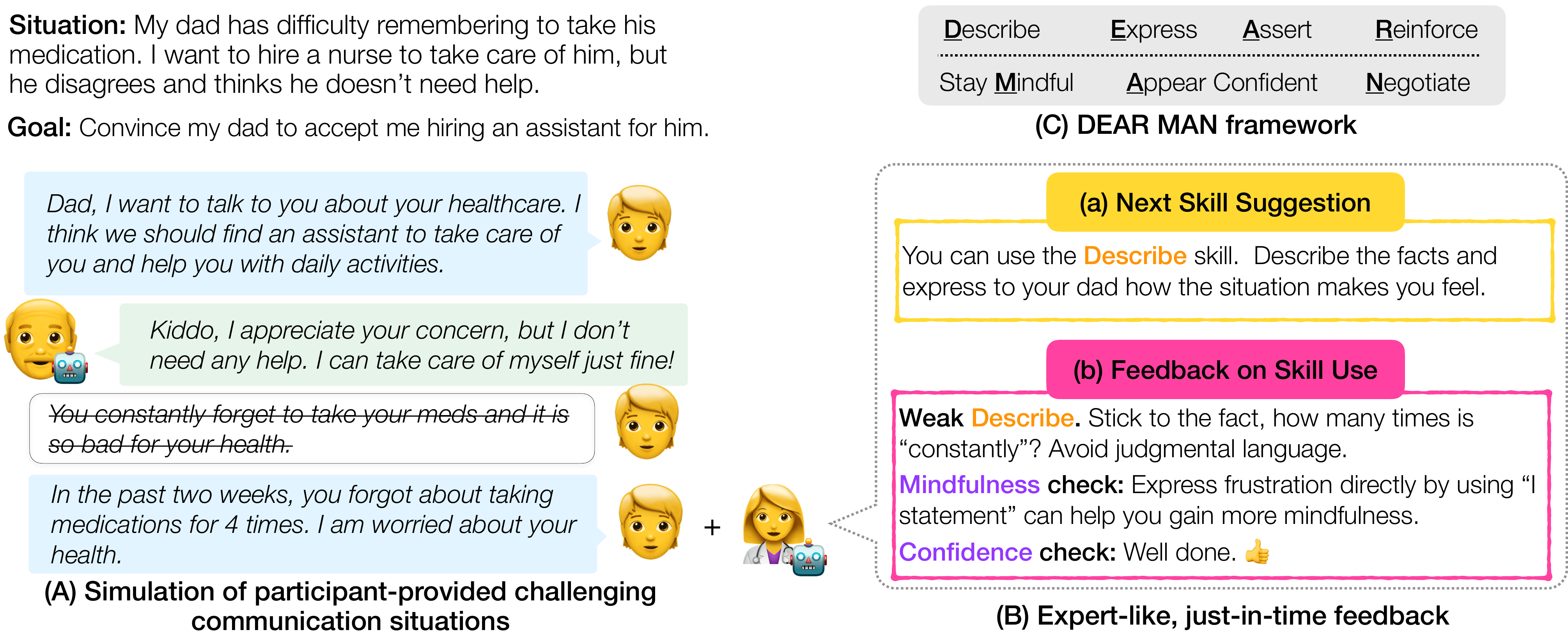}
    \vspace{-6mm}
    \caption{
    Overview of \ourmodel, an interactive training system that \textbf{(A)} simulates bespoke communication situations and \textbf{(B)} provides expert-like just-in-time feedback based on \textbf{(C)} the DEAR MAN framework. \ourmodels is backed by LMs that perform two tasks: \textbf{(a)} Next skill suggestion: before a user writes a message, \ourmodels suggests a skill to apply~(\S\ref{subsec:method-suggest-skill}). \textbf{(b)} Feedback on skill use: after a user writes a message, \ourmodels provides skill rating and improvement suggestions~(\S\ref{subsec:method-skill-use}).
    }
    \label{fig:figure_1}
    \vspace{-1.4em}
\end{figure*}

Some conversations can be challenging to navigate~\cite{stone2023difficult}, whether they concern negotiating a salary increase with a boss, discussing healthcare options with an aging parent, or asking a friend to return money they owe. Various communication frameworks assist individuals in conducting such conversations by providing a set of skills they can choose to apply~\cite{stone2023difficult, rosenberg2015nonviolent, linehan2014dbt, hartley2002interpersonal}.

However, psychology research highlights that a lack of communication skills is not the only obstacle to effective communication, particularly in emotionally charged situations~\cite{linehan2014dbt}. 
Difficult conversations can evoke strong emotions that disrupt effective communication, even for individuals with solid communication skills~\cite{LUFF20161461, rrepo19066}. To successfully communicate  during challenging situations, it is crucial to focus not only on communication skills, but also on managing emotions. 

The popular DEAR MAN framework, from Dialectical Behavioral Therapy (DBT), was 
originally developed for Borderline Personality Disorder, but is widely used to teach conversational strategies and emotional regulation~\cite{linehan2014dbt}. It includes conversational strategies (\textit{Describe, Express, Assert, Reinforce,} and \textit{Negotiate}) and a desired ``state of mind'' (\textit{Mindful} and \textit{Confident}) for productive conversations. Remaining mindful and confident in challenging conversations helps speakers regulate difficult emotions so they can successfully exercise their conversational strategies~(\S\ref{sec:framework};\ref{app:dearman-worksheet}).

Currently, DEAR MAN skills are mainly taught in therapy sessions and practiced either onsite in a roleplaying setting or at home with paper worksheets, which presents several challenges. Access to a trained therapist may be limited due to the significant shortage of mental health professionals~\cite{Olfson2016Shortage}. 
Outside of therapy sessions, static worksheets do not provide opportunities for interactive role playing and just-in-time feedback necessary for effective learning~\cite{beck1979cognitive,gagne1965analysis,beck1996beyond}.

Prior work in NLP has shown the ability of LMs to simulate personas and social interactions~\cite{argyle2023out, park2022social, park2023generative}. A few recent works leverage this capability by using LMs to help people improve interpersonal skills~\cite{liu2023improving} or conflict resolution skills~\cite{shaikh2023rehearsal}, without considering emotional regulation. 

Our work extends this literature, focusing on communication and emotional regulation skills simultaneously, incorporating expert domain knowledge into feedback, and grounding strategies in clinical psychology theory. 
We conduct a human-centered study and make three key contributions.


First, we present a formative study and an expert annotated dataset on DEAR MAN skill use. We conduct a formative study to gain insights from psychology experts on best practices when simulating challenging conversations and providing fine-grained feedback (\S \ref{sec:framework}). 
To understand how clinicians provide feedback on DEAR MAN in their practice and to develop and evaluate our method on real situations, we collect a dataset from crowdworkers consisting of difficult situations they encounter and simulated conversations within them (the crowdworker being paired with a role-playing LM partner).
We then ask psychology experts specifically trained in teaching DBT skills to annotate these conversations, assessing skill use and offering suggestions for improvement~(\S\ref{sec:data}). 


Second, we develop computational methods to provide feedback using insights from the formative study and collected dataset (\S\ref{sec:method}). We propose a new prompting strategy, demonstrating contrasting pairs of strong and weak utterances, in addition to state-of-the-art prompting methods. Our method improves the accuracy in skill use evaluation, outperforming GPT-4 by 24.8\%, and more expert-list, specific and actionable improvement suggestions. 

Third, we build \ourmodel, an interactive training system that \textit{simulates} difficult conversations and \textit{provide just-in-time feedback} backed by LMs to support the practice and learning of DEAR MAN skills~(Figure~\ref{fig:figure_1}). \ourmodels can be used at an individual's convenience to practice both communication and emotional regulation. Through a randomized trial with 86 participants, we evaluate \ourmodels's training outcomes on skill mastery, emotion reduction, and self-efficacy~(\S\ref{sec:user_study}). 
We show that a simulation-only variant of \ourmodels improves participants' self-efficacy towards having the conversation--boosting confidence (27\%) and reducing worries (4\%)--and emotion reduction towards the situations--reducing fear (16\%) and sadness (12\%)--while not improving skill mastery significantly. With the addition of just-in-time feedback, the participants' skill mastery significantly improved by 17.6\%, with additional improvement in self-efficacy (confidence, 26.7\%) and emotion reduction (fear, 15.7\%), compared to the simulation-only group.

%% file: 02_framework.tex
\section{Formative Study to Inform Design}
\label{sec:framework}
To understand how DEAR MAN skills are taught in practice, we conduct a formative study with clinical experts, summarizing crucial insights and corresponding design decisions below. Further details on the study procedure are in Appendix~\ref{app:formative_study}.

\textit{Insight 1: Guide clients to focus on facts instead of making judgemental comments when describing a situation.} We refrain from asking the participants to describe the personality of the conversation partner even though it may help LM simulate a more realistic conversation and instead only focus on past \textit{behaviors} that might influence the difficulty of the situation in \ourmodel.

\textit{Insight 2: 
Among the DEAR MAN skills, D, E, A, R, N are conversation strategies one can choose for each utterance. \textit{Mindful} and \textit{Confident} are the ``state-of-mind''. One should always stay mindful and confident throughout the entire conversation. } 
Therefore, each turn \ourmodels gives participants the option to choose from five conversation strategies. Conversation skills are evaluated only if they are selected for use, while mindfulness and confidence are assessed for each utterance~(\S\ref{sec:eval};\S\ref{sec:user_study}).

\textit{Insight 3: Practicing simpler or less emotionally intense situations helps with harder situations.} We collect difficulty levels from participants and use the less difficult situation in training~(\S\ref{sec:user_study}).

\textit{Insight 4: Training should prioritize emotion management. It is not considered a successful use of DEAR MAN skills if the client negotiates well but gets agitated.} We evaluate training outcomes in three aspects: skill mastery, emotion reduction, and self-efficacy (\S\ref{sec:user_study}). 

\textit{Insight 5: Choosing strategies before writing helps learning.} We adapt this design in data collection (\S\ref{sec:data}) and \ourmodel.


%% file: 03_data.tex
\section{Data Collection}
\label{sec:data}
We collect a dataset to understand how clinicians provide feedback on DEAR MAN in their practice and to develop and evaluate our method. 

\xhdr{Data collection with Crowdworkers}
We intend for our system to be used by individuals without specialized knowledge, so we collect data from crowdsourcing platforms. We recruit 20 people from Amazon mTurk who provided 60 different situations and annotations. Each worker provides three conversations, in each of family, social, work categories. Workers are asked to have conversations with our LM, which was instructed to roleplay as their partner during these conversations. Each conversation needs to be at least 10 responses from the worker, or until the simulated conversation partner ``agrees'' with the worker, whichever comes first. In each utterance, workers need to select one or more strategies they want to use in the given utterances to encourage them to follow the DEAR MAN framework as much as they could (\S\ref{sec:framework}). 
We include more ethics and safety details in \S\ref{sec:ethics};\ref{app:mturk}. 

\xhdr{DEAR MAN expert annotation}
We recruit six clinical experts have received specialized training and actively practice DBT. We only select those who indicate they ``sometimes'' or ``regularly'' work with clients on DEAR MAN skills on the sign up form~\S\ref{app:expert-recruitment-question}. Each expert annotated 2 to 4 conversations randomly selected from the dataset. In the final dataset, we have 18 conversations annotated, containing 163 utterances in total. For each utterance in a conversation, experts provided annotations on: 1) select the skills identified in the utterance, 2) rate the skill use with one of \texttt{strong} or \texttt{weak}\footnote{For mindful and confident, rate with \texttt{yes} or \texttt{no}.}, 3) for \texttt{weak} ratings, indicate suggestion for improvement and provide a rewritten utterance, 4) for skills not used, indicate reasons to use if the expert suggests to use them and provide a rewritten utterance. The interface for this annotation is shown in Appendix~\ref{appendix:sec:expert-interface}.

%% file: 04_method.tex
\begin{table*}[t!]
\small
\centering
\def\arraystretch{1.15}
\resizebox{\textwidth}{!}{%
\begin{tabular}{r|c|c|c|c|c|cccccccc}
\toprule
& \multicolumn{1}{c|}{\begin{tabular}[c]{@{}c@{}}Contrasting\\ Pairs\end{tabular}} & \begin{tabular}[c]{@{}c@{}}kNN\end{tabular} & \begin{tabular}[c]{@{}c@{}}Reas-\\-oning\end{tabular} & \begin{tabular}[c]{@{}c@{}}Curated \\ Rubric\end{tabular} & \textbf{Overall} & Describe                         & \multicolumn{1}{l}{Express}      & \multicolumn{1}{l}{Assert}       & \multicolumn{1}{l}{Reinforce}    & \multicolumn{1}{l}{Negotiate}    & \multicolumn{1}{l}{Mindful}      & \multicolumn{1}{l}{Confident}    \\ \midrule

\textbf{\ourmodels (Our method)} & \checkmark & \checkmark & \checkmark & \checkmark & \textbf{0.6442} & \textbf{0.7104} & 0.5797 & \textbf{0.6715} & \textbf{0.6873} & \textbf{0.7426} & 0.5965 & \textbf{0.5211} \\ \midrule
w/o Contrasting Pairs &  & \checkmark & \checkmark & \checkmark & 0.6248 & 0.6942 & \textbf{0.5847} & 0.6525 & 0.6257 & 0.7216 & \textbf{0.6159} & 0.4791 \\
w/o kNN (few-shot)&  &  & \checkmark & \checkmark & 0.5843 & 0.6220 & 0.5275 & 0.6425 & 0.5495 & 0.6651 & 0.6091 & 0.4757\\
w/o kNN (zero-shot) &  &  & \checkmark & \checkmark & 0.5756 & 0.5680 & 0.5797 & 0.5764 & 0.5900 & 0.6723 & 0.5381 & 0.5044 \\
w/o Reasoning  &  &  & & \checkmark & 0.5020 & 0.4552 & 0.5157 & 0.5427 & 0.5830 & 0.6651 & 0.5922 & 0.1602  \\
\midrule
GPT-4 &  &  &  &  & 0.3962 & 0.4690 & 0.4458 & 0.4340 & 0.4620 & 0.5018 & 0.3127 & 0.1480 \\ \bottomrule
\end{tabular}%
}
\vspace{-3mm}
\caption{Skill rating baseline and ablation results. We report macro F1 scores of binary classification of \texttt{Strong} vs not \texttt{Strong} use of each skill. \ourmodel, containing all four components: contrasting pair demonstrations, kNN demonstrations, reasoning step, and curated rubric, achieves the highest macro F1 overall, with significant outperformance on Describe, Assert, Reinforce, Negotiate, and Confident skills. \ourmodels outperforms GPT-4 by 24.8\%.}
\label{tab:feedback-eval}
\vspace{-1em}
\end{table*}
\section{Methodology}
\label{sec:method}

\ourmodels is an interactive training system that simulates bespoke communication situations and provides expert-like just-in-time feedback based on the DEAR MAN framework. \ourmodels is backed by LMs that perform two tasks: (a) Next skill suggestion: before a user writes a message, \ourmodels suggests a skill to apply~(\S\ref{subsec:method-suggest-skill}), (b) Feedback on skill use: after a user writes a message, \ourmodels provides skill rating and improvement suggestions~(\S\ref{subsec:method-skill-use}). We describe our methods for performing these tasks.

\subsection{Skill rating and improvements suggestions}
\label{subsec:method-skill-use}
To ensure low-latency and cost-efficiency, we define a multitask problem: for a situation $S$, an utterance $U_i$, and a skill $L_i$, simultaneously generate skill rating $R_i$ and improvement suggestions $F_i$. 

The major challenges include operationalizing complex DEAR MAN constructs grounded in psychology and supporting the variety of situations users may want to simulate. 
Previous research has shown the effectiveness of in-context learning for various NLP tasks~\cite{Brown2020LanguageMA, sharma2023cognitive}. 
Our method builds on these approaches with four key components: 1) curated rubrics to augment the LMs with experts' insights in~\S\ref{sec:data}, 2) a reasoning step for both demonstrations and generation to facilitate skill rating, 3) kNN retrieval of few-shot demonstrations from the expert-annotated data in~\S\ref{sec:data}, and 4) contrasting pair demonstrations to help LMs learn nuanced concepts. 

\xhdr{Curated rubric} To enhance the model's rating calibration, we incorporate information extracted from expert-written feedback into the rating rubric. We use DBSCAN~\cite{ester1996density} to cluster feedback on \texttt{weak} ratings and where a skill should be applied but was not. Through qualitative evaluation, we tune parameters in order to identify one distinct reason per cluster. We then summarize common reasons for each skill and integrate them into the system prompts (Appendix ~\ref{app:system_prompts}).

\xhdr{Reasoning step} We follow previous work using chain-of-thought prompting~\cite{Wei2022ChainOT} to generate the reasoning of a rating before assigning it. We convert expert-written suggestions into reasoning of ratings and use them as demonstrations. e.g., a suggestion ``don’t mix feelings and facts'' is converted into a reason ``the utterance mixes feelings and facts.'' We perform the conversion using few-shot learning and qualitatively evaluate the conversion with a random sample of 50. 

\xhdr{kNN demonstrations} Retrieval-based in-context learning has shown superior performance to comparable approaches in similar tasks~\cite{sharma2023cognitive}. We adapt this approach and retrieve a set of examples from all levels of skill use. We first encode all utterances using the \texttt{all-mpnet-base-v2} model with SentenceTransformer. For each query utterance, we use \texttt{faiss}~\cite{douze2024faiss} to retrieve the $k$ most similar examples from each level (\texttt{strong}, \texttt{weak}, \texttt{none}) for this skill in our datasets.

\xhdr{Contrasting pair demonstrations} Utterances often involve the use of multiple skills, posing a challenge for models to identify the text corresponding to each skill. 
To address this challenge, we construct pairs of (\texttt{strong}, \texttt{weak}) and (\texttt{strong}, \texttt{none}) demonstrations. We first search for the $k$ \texttt{weak} and \texttt{none} examples that are most relevant to the query utterance. 
We then use the expert rewritten responses as \texttt{strong} examples to form the contrasting pairs, and use these pairs as demonstrations, which helps the model learn nuanced concepts and disentangle multiple skills. 
For instance, in the utterance: ``\ul{\textit{In recent team meetings, my ideas were presented as yours }}(Strong Describe)... \textit{this situation has been causing some discomfort} (Strong Express).'' Without contrasting pair demonstrations, a model misclassifies it as Weak Describe, suggesting a mixture of facts and feelings. Classifying skill use as weak would trigger unnecessary, if not confusing feedback. However, the underlined \ul{subspan} corresponding to Describe remains focused on facts, qualifying it as a Strong Describe. We demonstrate empirically that a contrasting pair prompting strategy improves skill rating prediction and the quality of improvement suggestions~(\S\ref{sec:eval}).

\subsection{Next skill suggestion}
\label{subsec:method-suggest-skill}
Before a participant writes utterance $U_i$, we aim to suggest the set of best skills to use, given situation $S$ and previous simulated partner's response $P_{i-1}$. In our dataset, skill $L_j$ is considered ``recommended'' if: 1) $L_j$ is selected by the participant and the expert does not advise against it, or 2) $L_j$ is not selected but is suggested by the expert. 
Based on insights from~\S\ref{sec:framework}, we design the model to always suggest \textit{describe} as the first skill (when $i=0$). For $i>=1$, we retrieve the $k$ most similar examples to $S$ concatenated with $P_{i-1}$, to prompt GPT-4 and generate the suggested skill.

%% file: 05_experiments.tex
\begin{table}[]
\small
\centering
\def\arraystretch{1.15}
\resizebox{\columnwidth}{!}{%
\begin{tabular}{r|c|cc|cc}
\toprule
 & Human & R-L & BScore & Spec. & Act. \\ \midrule
\textbf{\ourmodels} & \textbf{83\%} & {10.3} & \textbf{85.2} & \textbf{4.05} & \textbf{4.04}  \\
\midrule
w/o Contrasting Pair & 51\% & \textbf{11.2} & {85.0} & 4.01 & 3.87 \\
w/o kNN (few-shot) & 23\% & 9.0 & 84.3 & 2.55 & 2.38 \\
w/o kNN (zero-shot) & 37\% & 8.7 & 84.0 & 3.75 & 2.54 \\
w/o Reasoning & 34\% & 7.7  &  83.8 & 3.18 & 2.51 \\
\midrule
GPT-4 & 6\% & 5.8 & 81.0 & 1.25 & 2.28  \\
\midrule
DEAR MAN Experts & 100\% & 100 & 100 & 3.02 & 4.01 \\
\bottomrule
\end{tabular}%
}
\vspace{-3mm}
\caption{Similarity between generated vs experts' improvement suggestions. \ourmodels achieves competitive R-L and BScore and the best human evaluation performance. 83\% of the time, \ourmodels is essentially suggesting the same improvements as DEARMAN experts, based on human eval. Note that automatic metrics should be interpreted with caution, as the gaps in human evaluations are significantly larger. \ourmodels achieves highest \textit{specificity} and \textit{actionability}, providing specific and actionable improvement suggestions even more than expert-written suggestions. R-L: ROUGE-L; BScore: BertScore; Spec.: Specificity; Act.: Actionability} 

\label{tab:suggestion_eval}
\end{table}

\section{Evaluation of \ourmodels with an Expert-Annotated Dataset} 
\label{sec:eval}
We evaluate \ourmodels on the expert-annotated dataset (\S\ref{sec:data}) with cross-validation. 
We use \ourmodels to generate skill use feedback (\S\ref{subsec:method-skill-use}) and next skill suggestions (\S\ref{subsec:method-suggest-skill}) and report the average performance across all conversations\footnote{To ensure deterministic skill rating predictions, we use \texttt{gpt-4-1106-preview} with \texttt{temp=0} throughout this paper. }.

Since there are no established methods for the proposed new tasks, we use GPT-4 as a baseline and conduct the following ablations to assess the impact of each component in \ourmodel~(Tables~\ref{tab:feedback-eval}\&\ref{tab:suggestion_eval}): (1) without contrasting pairs component (retrieval-based in-context learning, with reasoning step and curated rubric), (2) without contrasting pairs or kNN-retrieval component (random in-context examples), with reasoning step and curated rubric (3) without in-context examples, only reasoning step and curated rubric, (4) without in-context examples or reasoning, only curated rubrics.

\xhdr{Skill ratings} To maximize feedback opportunities, we prioritize identifying the distinction between \texttt{strong} vs. not \texttt{strong} skill use, which determines whether the model will provide improvement suggestions. As shown in Table~\ref{tab:feedback-eval}, \ourmodels achieves the highest macro F1 on average across skills, outperforming GPT-4 by 24.8\%. \ourmodels outperforms GPT-4 baseline and all ablations on five out of seven DEAR MAN skills. 

\begin{table}[]
\small
\centering
\def\arraystretch{1.15}
\resizebox{\columnwidth}{!}{%
\begin{tabular}{l|cc}
\toprule
 & \textbf{Macro F1} & \textbf{Entropy} \\ \midrule
\textbf{kNN few-shot} & \textbf{0.5849} & {\ul{2.23}} \\ \midrule
Random few-shot & {\ul{0.5723}}& 1.55 \\
Zero-shot & {0.5345}& 1.59 \\
Always suggest the most frequent (Assert) & 0.4780 & 0.00 \\
Random suggestion & 0.3899 & \textbf{2.29} \\ \bottomrule
\end{tabular}%
}
\vspace{-3mm}
\caption{Next skill suggestion, evaluation with expert-annotated dataset. \ourmodels gives diverse skill recommendations, almost retaining max entropy (uniform random suggestion) while improving 9\% over the \underline{second best} method's F1 score without kNN demonstrations.}
\label{tab:skill-suggestion-results}
\vspace{-15pt}
\end{table}

\begin{figure*}
    \centering
    \includegraphics[width=\textwidth]{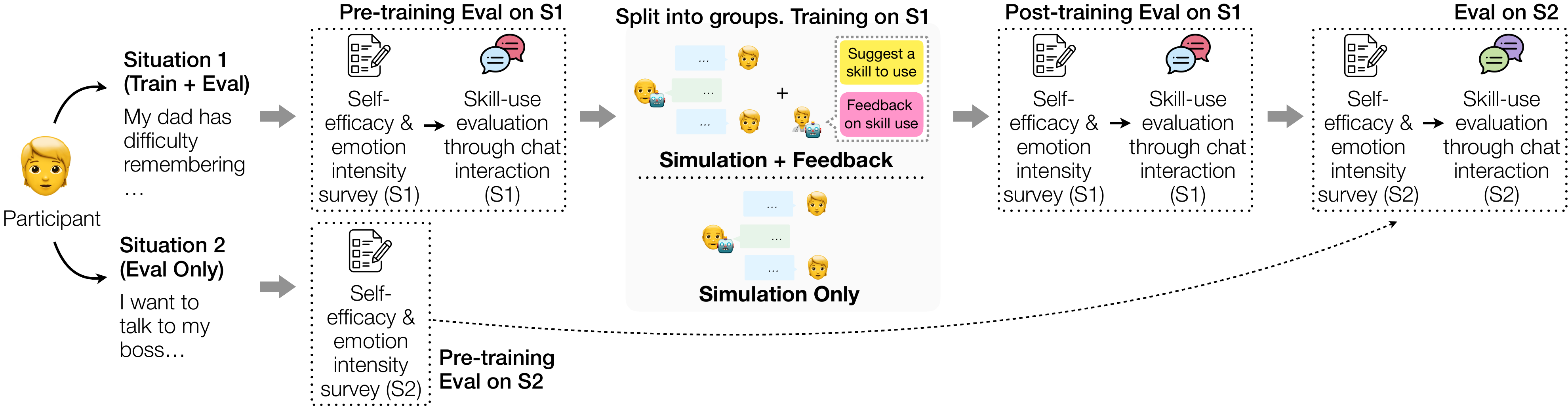}
    \vspace{-2em}
    \caption{User study experimental design. We randomly assigned participants to one of simulation-only and simulation+feedback groups. Each participant was asked to provide two situations, S1 and S2. Only S1 was used in training. Both S1 and S2 were used in pre- and post-training self-efficacy and emotion intensity surveys and in post-training skill-use evaluation through chat interaction.}
    \label{fig:user-study-procedure}
    \vspace{-1em}
\end{figure*}

\xhdr{Improvement suggestions} We compare the generated improvement suggestions with expert-written suggestions through a combination of human and automatic evaluation. In our human evaluation, we recruit CS PhD students with significant expertise in NLP and text annotation tasks and ask them to annotate if the expert and model-generated improvement suggestions are similar (on a random sample of 210 pairs; details in~\S\ref{app:human_eval}). We find that \ourmodels significantly outperforms baseline and ablations, generating improvement suggestions similar to experts 83\% of the time, which is 32\% better than the second best. Moreover, we conduct a secondary evaluation using automatic metrics, ROUGE-L~\cite{lin2004rouge} and BertScore~\cite{zhang2019bertscore}, and find that our model is competitive on both metrics, compared with baseline and ablations. Note that given the open-ended nature of improvement suggestions, automatic metrics are often limited in their ability to capture nuances in what should be considered similar, often focusing on the semantic and linguistic similarity instead of the similarity of the underlying feedback.

We also evaluate the \textit{specificity} and \textit{actionability} of the improvement suggestions. Prior work in NLP to support mental health skills has suggested that \textit{specific} and \textit{actionable} feedback is highly preferred and more effective ~\cite{sharma2023cognitive}. Here, we use a simple GPT-4-based few-shot prompting method \cite{ziems2023can} to measure specificity and actionability. \ourmodels outperforms baseline and ablations in both measures, even more than experts, who might be too busy to consistently write highly specific feedback. \ourmodels is comparable with experts in \textit{actionablity}, significantly outperforming all baseline and ablations.

\xhdr{Next skill suggestion performance and diversity} To ensure that users receive a diverse range of skill suggestions for practice, we evaluate both the performance of predicting "expert-recommended" skills, and the diversity of the skill suggestions through entropy. As Table~\ref{tab:skill-suggestion-results} shows, our method surpasses the second-best baseline by 9.4\% in performance with almost maximum entropy\footnote{\ourmodels is able to recommend multiple skills at the same time, here we evaluate on single skill recommendation, so users can focus on improving one skill at a time.}.

%% file: 06_user_study.tex

\section{Evaluate \ourmodels in a Randomized Trial}
\label{sec:user_study}

We conduct a randomized trial with 86 participants and assess how \ourmodels can help people improve interpersonal effectiveness. 

\subsection{Participant Training Methods}

We evaluate two variants for training participants on interpersonal effectiveness -- (1) Simulation Only (S) and (2) Simulation + Feedback (S + F). 

\xhdr{(1) Simulation Only (S)}
We develop a GPT-4-based role-playing chatbot designed for participants to converse about their situation (e.g., a chatbot role-playing as the participant's boss). The role-playing chatbot leverages the situation to create a system prompt for GPT-4 (\S\ref{app:system_prompt_simulation}). Also, it is designed to be difficult to convince and respond at lengths similar to the length of the participant's message. We qualitatively evaluate this chatbot during our formative study (\S\ref{sec:framework}) and data collection (\S\ref{sec:data}). Participants interact with the chatbot to \textit{simulate} the conversation.

\xhdr{(2) Simulation + Feedback (S+F)} Using the model developed in  \S\ref{sec:method}, we generate the following types of interactive feedback for participants: (1) get a skill suggestion (\S\ref{subsec:method-suggest-skill}), (2) select a skill (can be different from what is suggested) and write a response implementing this skill, (3) get feedback (ratings + improvement suggestions) on skill use (\S\ref{subsec:method-skill-use}), (4) improve the response based on the feedback. Steps (2)-(4) can be optionally repeated. Participants receive this feedback, while interacting with the role-playing chatbot designed above to simulate the situation.
 
To compare \ourmodels with current at-home practice, participants in both variants get the DEAR MAN worksheet from the official DBT manual~\cite{linehan2014dbt}, mirroring current practice.

\subsection{Study Procedure and Evaluation Metrics}
Figure~\ref{fig:user-study-procedure} outlines our study procedure. We recruit participants from mTurk ($n=34$) and Prolific ($n=52$). Each participant is asked to provide two difficult communication situations (S1 and S2). Next, they are randomly assigned either Simulation Only (S) or Simulation Feedback (S+F). More details about study interface procedure are in~\S\ref{app:user-study-interface}.

The DEAR MAN manual suggests that people's primary struggles in challenging situations are: lack of skills, interference of strong emotions, and fear of not having a successful conversation~\cite{linehan2014dbt}. Here, we measure the improvement in DEAR MAN \textbf{skill mastery}, \textbf{emotion reduction} towards the challenging situation, and \textbf{self-efficacy} towards having these conversations. We evaluate them \textit{pre-} and \textit{post-}training, enabling a within-person control setup. We also compare the differences between the S and S+F groups, distilling the effect of just-in-time feedback.

\begin{figure}[t!]
\includegraphics[width=.95\columnwidth]{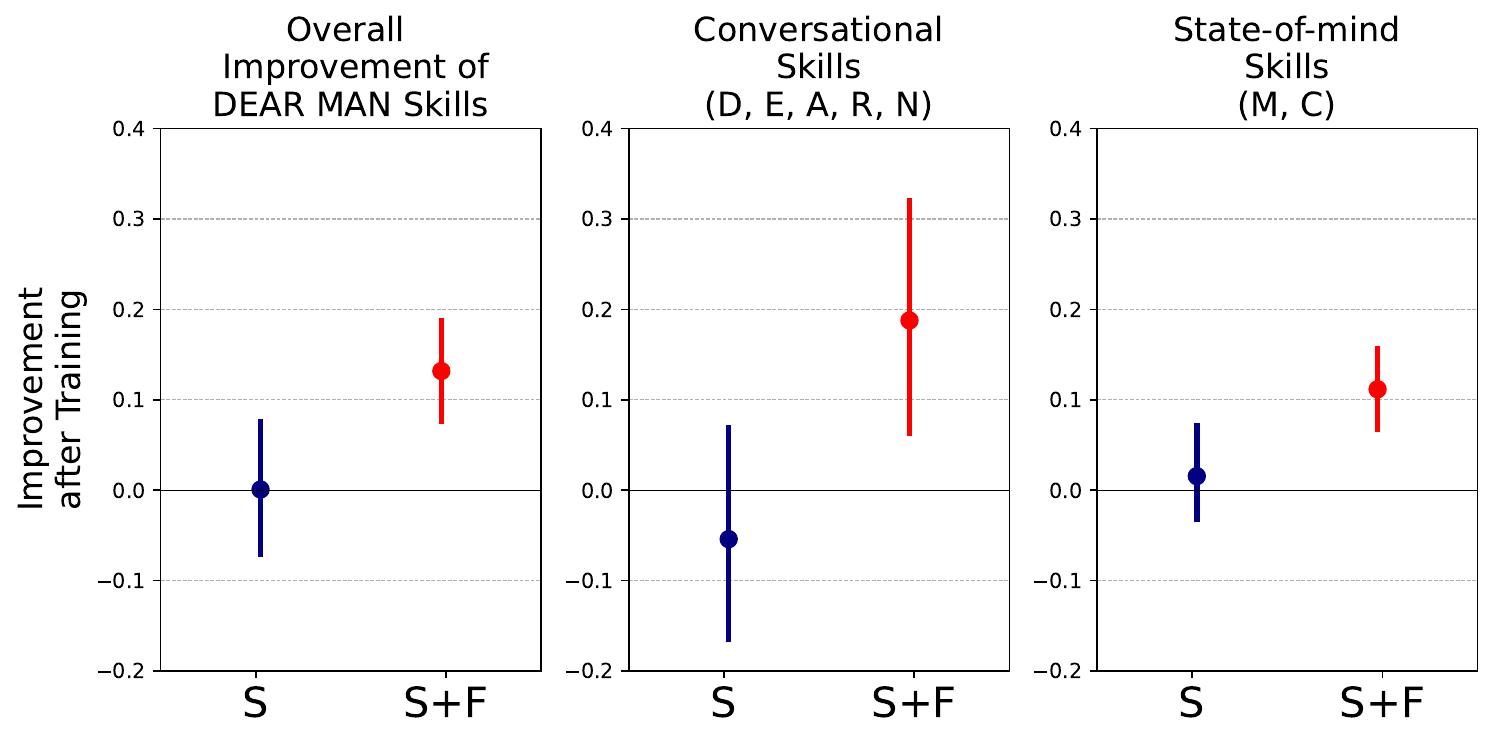}
\vspace{-2mm}
\caption{Improvement in skill mastery. Simulation+feedback group shows a significantly higher improvement in skill mastery (17.6\% on a 0-2 scale, **,  $d$=0.59) compared to simulation-only (0.1\%) after only one training session. The difference is also significant for the subset of conversational skills that participants choose to use in each utterance (only measured when the skills are chosen), \textit{Describe}, \textit{Express}, \textit{Assert}, \textit{Reinforce}, and \textit{Negotiate} (24.8\%,  **, $d$=0.59) and state-of-mind skills (measured in every utterance), \textit{Mindful} and \textit{Confident} (15.7\%, **, $d$=0.59). (***: $p<.001$, **:$p<.01$, *:$p<.05$. $d$: Cohen's $d$.)}
\label{tab:user-study-results-skill-mastery}
\vspace{-1.5em}
\end{figure}

\begin{figure*}[t!]
\includegraphics[width=\textwidth]{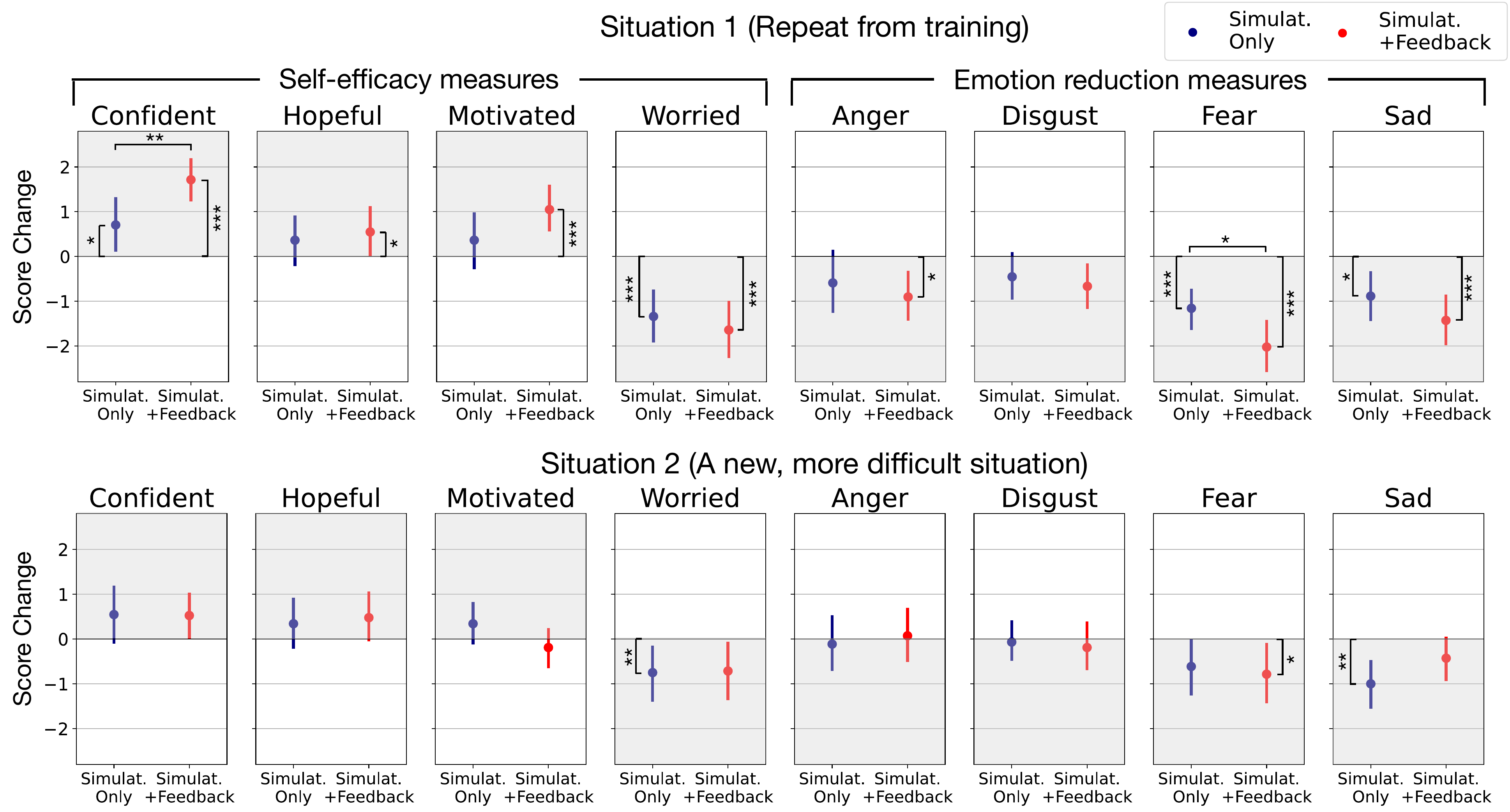}
\vspace{-8mm}
\caption{Change of self-reported efficacy and emotional intensity for both Situation 1 (S1) and Situation 2 (S2) after a single training session on S1. \colorbox{lightgray}{Gray area} indicates the direction of improvement for each score. The group receiving just-in-time feedback generated with our method in addition to conversation simulation see significant increase in their confidence (43.6\%, ***), hopefulness (11.0\%, *), motivation (22.1\%, ***) towards having the conversation, significant decrease in their worrying thoughts (30.9\%, ***) about having the conversation and their anger (23.5\%, *), fear (40.9\%, ***), and sadness (29.0\%, ***) towards the training situation (S1). The increase in confidence and reduction in fear are 26.7\% (**, $d$=0.57) and 15.7\% (*,$d$=0.51) significantly more than the group receiving simulation only. This improvement in self-efficacy and emotional reduction does not transfer immediately to a new, more difficult situation (S2). See Section \ref{sec:user_study} for more analysis and discussion.  }
\vspace{-1em}
\label{fig:self-reported-results}
\end{figure*} 
\subsection{RQ1: How do simulation and feedback improve DEAR MAN skill mastery?} We measure a participant's skill mastery with our model in \S\ref{sec:method} before and after the training. We then compare the level of skill use in\textit{ pre-} and \textit{ post-}training evaluation chat to evaluate the effect of training on the situation that the particiant is being trained on (S1). Also, to test the generalizability of the skill learning, we also conduct evaluation on a new and more difficult situation (S2) in which the participant has not been trained on and does not receive feedback during conversation simulation.

Figure~\ref{tab:user-study-results-skill-mastery} compares S and S+F groups on the improvement of skill mastery for S1, the situation also used during training. S+F group shows a significantly higher improvement in skill mastery after one training conversation by 17.6\% ($p = .007$, Cohen's $d = 0.59$) on a score scale of 0-2, compared to the S group which improved by only 0.1\%. This difference is also significant for the set of five conversation skills, \textit{Describe}, \textit{Express}, \textit{Assert}, \textit{Reinforce}, and \textit{Negotiate}, at 24.8\% ($p=.008, d=0.59$) and the set of state-of-mind skills, \textit{Mindful} and \textit{Confident}, at 15.7\% ($p=.007, d=0.59$). Among all the skills, S+F shows significant more improvement in \textit{Express}, \textit{Mindful}, and \textit{Confident} (Appendix Figure~\ref{app:utterance-level-comparison}).

\subsection{RQ2: How do simulation and feedback enhance emotion reduction?} We evaluate emotion reduction on four negative emotions from the Plutchik’s Wheel \cite{Plutchik1980AGP}: anger, fear, sadness, disgust. We ask the participants to rate their agreement to statements like ``I feel sad about the situation.’’ use a 7-point Likert scale~\cite{likert1932technique} before and after the training. 
S+F group shows significant reduction of almost all negative emotions on S1. We find that both S group and S+F group have reduced \textbf{fear} (by 25.1\%, 40.8\%, $p=.000,.000$, $d=.71, 1.19$) and \textbf{sadness} (by 17.3\%, 29.9\%, $p=.020, .000$, $d=.45, .76$) towards the situation after training. S+F group shows a significantly higher reduction towards \textbf{fear} (by 15.7\%, $p=.021$, $d=.51$), compared to S group. S+F group also has a significant reduction in \textbf{anger} (by 23.5\%, $p=0.030$), whereas the S group does not show significant change. 

\subsection{RQ3: How do simulation and feedback improve participants' self-efficacy?} To evaluate participants' self-efficacy before and after the training, we ask the participants to rate their confidence, worry, hopefulness, and motivation about having the conversation before and after the training, again with a 7-point Likert-scale. 

As Figure~\ref{fig:self-reported-results} shows, both S and S+F groups show a significant increase in self-reported \textbf{confidence} (by 16.9\%, 43.6\%, $p=.035, .000$, $d=.41, 1.08$) and a significant reduction in self-reported \textbf{worry} (by 26.9\%, 30.9\%, $p=.000,.000$, $d=.81,1.04$) towards having the conversation.
Moreover, S+F group demonstrates significantly higher increases in \textbf{confidence} (by 26.7\%, $p=.010$, $d=.57$), compared to the S group, underscoring the effectiveness of just-in-time feedback.
In addition, S+F group showed a significant increase in \textbf{hopefulness} (by 11.0\%, $p=.046$, $d=.35$) and \textbf{motivation} (by 22.1\%, $p=.001$, $d=.62$) towards having the conversation, whereas the S group does not show significant change. This shows that S+F version of the tool helps in these dimensions, though we cannot separate the effect of just-in-time feedback and repeated practice with the simulation. 


\subsection{RQ4: Do these effects generalize to a new and more difficult situation?}
We compare the average skill used in conversation on S2 and both \textit{pre-} and \textit{post-}training evaluation conversation on S1. We find that for S+F group, the average skill use rating is significantly higher in conversation on S2 compared to \textit{pre-}training conversation on S1 ($p=.049$). The skill use ratings are not significantly different between conversation on S2 and \textit{post-}training conversation on S1. These comparisons show that the skill use improvement can be generalized, without significant diminishing effect, to a new and more difficult situation immediately after training. 

Although skill mastery generalizes to a new and more difficult situation (S2), self-efficacy and emotional reduction do not immediately generalize~(Figure~\ref{fig:self-reported-results}). Many constructs, such as confidence, hopeful, worry, fear and sadness, show a positive improvement but these differences were not statistically significant at $\alpha = 0.05$. This could be attributed to the difficulty of managing emotions in novel situations without specific training, suggesting that targeted emotional regulation training of a different type or over an extended period may be necessary~\cite{freitas2000regulating}. 

The findings also emphasize that practicing in simulations with feedback tailored to the exact situation is more effective for improving self-efficacy and managing emotions. Our tool supports exactly this accessibility, lowering the barrier to effective learning and practices.

%% file: 07_related_work.tex
\section{Related Work}
\label{sec:related_work}
Broadly, our work is related to the growing body of works on LLM-based autonomous agents~\cite{park2022social, park2023generative,argyle2023out,argyle2023leveraging,Zhou2023SOTOPIAIE,Liu2023TrainingSA,Aher2022UsingLL,Wang2023ASO,Dubois2023AlpacaFarmAS} and using LLM in psychology and computational social science~\cite{Ziems2023CanLL,Demszky2023UsingLL,sharma2023cognitive,sharma2023,lin-etal-2022-gendered,perez2022pair,shah2022modeling,sharma2020engagement,sharma2020computational,wadden2020effect,welch2020expressive,zhang2020balancing,gaur2019knowledge,lee2019identifying,perez2019makes,althoff2016large}. Our work most closely relates to recent works using LMs in their roleplaying capacity to facilitate communication skill learning~\cite{shaikh2023rehearsal, liu2023improving, argyle2023leveraging}. Our work is the first to focus on both communication skills and emotion management simultaneously, incorporate experts' domain knowledge in providing feedback, and ground in clinical psychology theory.

%% file: 08_conclusion.tex
\section{Conclusion}
\label{sec:conclusion}
In this paper, we demonstrate how Human-LM interaction can be used to facilitate interpersonal effectiveness skill learning and practice. We collect a dataset with crowdworkers and clinical experts who are specifically trained and in practice of DBT. Using this dataset, we develop methods to prompt LMs to simulate bespoke communication scenarios and provide just-in-time feedback, grounded in psychotherapy theory. We build an interactive training system \ourmodel, and conduct a randomized user study with 86 participants to assess the effectiveness of the simulation and feedback components of \ourmodel. We find that simulation-only training is effective in improving self-efficacy and emotion reduction, and adding just-in-time feedback shows significantly more benefits in all of skill mastery, self-efficacy, and emotion reduction. The skill mastery can be acquired from practicing with a different situation, while emotion reduction and self-efficacy appear to only benefit from training specifically on the situation.

%% file: 09_limitations_ethics_statement.tex

\section{Ethics Statement }
\label{sec:ethics}

\xhdr{IRB Approval} We obtained approval from our institution’s Institutional Review Board (IRB) for our study. Our institution requires all researchers who conduct human subjects research to complete human subjects protection training. The researchers conducted this study were certified by IRB.

\xhdr{Informed Consent from Participants} We obtained consent from participants in both our data collection and the user study. All participants are aged 18 and older. Participants were informed that they were interacting with an AI-based model simulating their conversation partner and the data they provided will be released for research purposes. Participants were also informed that some content from the model might be upsetting since the conversation might get heated. 

\xhdr{Crisis Resources} We use an API with content filters to minimize the possibility of harmful output during deployment. \footnote{https://learn.microsoft.com/en-us/azure/ai-services/openai/concepts/content-filter} Nevertheless, some content might still be upsetting to the participants. We provide two crisis resources to our participants during the study: Crisis Text Line (crisistextline.org) and 988 Suicide and Crisis Lifeline (988lifeline.org). We did not observe any adverse events.

\xhdr{Privacy} Our study does not collect Privately Identifiable Information (PII) and we asked that participants avoid including any PII in the situations or conversations. The conversations and situations were manually filtered to ensure there were no identifiable names or locations. 


\section{Limitations} 

Our work is not without limitations. Importantly, we note that our tool is not meant to replace practice with an expert. Rather, we built the tool to complement current practice and to lower the barrier of access to learning and practicing. We note further limitations below.

We do not provide best negotiation strategies but rather focus on the wellbeing and mindfulness of the conversation participant. For example, we consider it a suboptimal case if someone ``wins'' a negotiation but was not being mindful and had negative emotional swings during the process, this is based on insights from experts in~\S\ref{sec:framework}. 

We do not assist participants in setting their goals. In our randomized trial, we choose participants who can clearly express their goals and work with them to achieve these goals. Setting the right goal is crucial but can be challenging, and other frameworks in DBT address this issue. We have begun by collecting goals and expert annotations in our data collection for future research to expand upon.

Due to the design of our study, the study length is already about an hour. To avoid cognitively overloading the participants, we asked them to do only one training session. We did not investigate the effect of different ``dosages'' of training. In addition, short-term improvement may not imply long-term improvement, further work is needed to investigate the long-term effect of using such a tool. However, we note that a key benefit of our system is the just-in-time availability that allows practice just before the user anticipates a challenging conversation. 

To minimize participant burden, we collect self-reported scores for emotion reduction and self-efficacy constructs through single questions, rather than a comprehensive survey. We use common measures like "sad" and "angry" for emotions and like "confident" and "worry" for self-efficacy to prevent reporting biases due to misinterpretation. However, it is important to note that self-reported scores, while commonly used in mental health assessments, may contain biases and inaccuracies~\cite{stone1999science}.

Our experimental design does not consider individuals with specific mental health conditions that could impact communication. Additionally, we do not address cultural variations in communication, recognizing that what may be perceived as confidence in one culture could be seen as aggression in another. We leave it to future work to develop more personalized and culturally sensitive communication training tools.

Language models are known to contain biases~\cite{Santurkar2023WhoseOD, Zhou2023NavigatingTG, Durmus2023TowardsMT, lin-etal-2022-gendered,aguirre2023selecting}. In our context, the simulation step may contain persona bias~\cite{gupta2024personabias}. Our tool, designed with \textit{Insight 1} in \S\ref{sec:framework}, steers participants to focus on facts and avoids characterizing personalities. This mitigates the risk of triggering the LM to exhibit persona biases. Nonetheless, thorough assessment of bias and safety is necessary prior to deploying a tool of this nature in the real world.


%% file: 10_Appendix.tex
\onecolumn
\appendix
\input{100_App_02formative}

\newpage
\section{User Study Results }

\begin{table}[h]
\resizebox{\textwidth}{!}{%
\begin{tabular}{@{}llllllll@{}}
\toprule
 & treatment\_post\_scores & treatment\_pre\_scores & perc\_t & treatment\_t\_stat & treatment\_p\_value & symbol & Cohen's d \\ \midrule
Confident & 5.64 & 3.93 & 44.0\% & 6.48 & 0.00 & *** & 1.08 \\
Worried & 3.67 & 5.31 & -31.0\% & -5.2 & 0.00 & *** & -1.04 \\
Hopeful & 5.52 & 4.98 & 11.0\% & 2.05 & 0.05 & * & 0.35 \\
Motivated & 5.79 & 4.74 & 22.0\% & 3.68 & 0.00 & *** & 0.62 \\
Anger & 2.95 & 3.86 & -23.0\% & -2.25 & 0.03 & * & -0.47 \\
Fear & 2.93 & 4.95 & -41.0\% & -6.54 & 0.00 & *** & -1.19 \\
Disgust & 2.38 & 3.05 & -22.0\% & -1.85 & 0.07 &  & -0.33 \\
Sad & 3.5 & 4.93 & -29.0\% & -4.77 & 0.00 & *** & -0.76 \\ \bottomrule
\end{tabular}%
}
\caption{User Study Results - Simulation+Feedback. \textbf{Situation 1}. Improvement after the training for Treatment and Control groups. Significance means there is a significant increase of self-reported efficacy or emotions after training. }
\label{tab:app-user-study-TC-survey-S1}
\end{table}

\begin{table}[h]
\resizebox{\textwidth}{!}{%
\begin{tabular}{@{}llllllll@{}}
\toprule
 & control\_post\_scores & control\_pre\_scores & perc\_c & control\_t\_stat & control\_p\_value & symbol\_c & Cohen's d \\ \midrule
Confident & 4.86 & 416.0\% & 0.17 & 2.18 & 0.03 & * & 0.41 \\
Worried & 3.64 & 498.0\% & -0.27 & -4.5 & 0.00 & *** & -0.81 \\
Hopeful & 4.91 & 455.0\% & 0.08 & 1.21 & 0.23 &  & 0.2 \\
Motivated & 5.09 & 473.0\% & 0.08 & 1.12 & 0.27 &  & 0.2 \\
Anger & 3.61 & 420.0\% & -0.14 & -1.53 & 0.13 &  & -0.32 \\
Fear & 3.45 & 461.0\% & -0.25 & -3.87 & 0.00 & *** & -0.71 \\
Disgust & 2.7 & 316.0\% & -0.14 & -1.42 & 0.16 &  & -0.27 \\
Sad & 4.25 & 514.0\% & -0.17 & -2.42 & 0.02 & * & -0.45 \\ \bottomrule
\end{tabular}%
}
\caption{User Study Results - Simulation-Only. \textbf{Situation 1}. Improvement after the training for Treatment and Control groups. Significance means there is a significant increase of self-reported efficacy or emotions after training. }
\label{tab:app-user-study-TC-survey-S1}
\end{table}

\begin{table}[h]

\resizebox{\textwidth}{!}{%
\begin{tabular}{@{}lllllll@{}}
\toprule
emotion & treatment\_post\_scores & treatment\_pre\_scores & perc\_t & treatment\_t\_stat & treatment\_p\_value & symbol \\ \midrule
Confident & 4.214 & 3.690 & 0.142 & 1.834 & 0.074 &  \\
Worried & 4.714 & 5.429 & -0.132 & -1.888 & 0.066 &  \\
Hopeful & 4.905 & 4.429 & 0.108 & 1.800 & 0.079 &  \\
Motivated & 4.881 & 5.071 & -0.038 & -0.840 & 0.406 &  \\
Anger & 4.548 & 4.476 & 0.016 & 0.215 & 0.831 &  \\
Fear & 4.214 & 5.000 & -0.157 & -2.118 & 0.040 & * \\
Disgust & 3.357 & 3.548 & -0.054 & -0.797 & 0.430 &  \\
Sad & 4.548 & 4.976 & -0.086 & -1.232 & 0.225 &  \\ \bottomrule
\end{tabular}%
}
\caption{User Study Results - Simulation+Feedback. \textbf{Situation 2.} Improvement after the training for Treatment and Control groups. Significance means there is a significant increase of self-reported efficacy or emotions after training.}
\label{tab:app-user-study-TC-survey-S2}
\end{table}

\begin{table}[h]

\resizebox{\textwidth}{!}{%
\begin{tabular}{@{}lllllll@{}}
\toprule
 & control\_post\_scores & control\_pre\_scores & perc\_c & control\_t\_stat & control\_p\_value & symbol\_c \\ \midrule
Confident & 4.068 & 3.523 & 0.155 & 1.312 & 0.196 &  \\
Worried & 4.545 & 5.295 & -0.142 & -2.096 & 0.042 & * \\
Hopeful & 4.341 & 4.000 & 0.085 & 1.106 & 0.275 &  \\
Motivated & 4.682 & 4.341 & 0.079 & 0.965 & 0.340 &  \\
Anger & 3.955 & 4.068 & -0.028 & -0.292 & 0.772 &  \\
Fear & 4.273 & 4.886 & -0.126 & -1.414 & 0.165 &  \\
Disgust & 3.250 & 3.318 & -0.021 & -0.230 & 0.819 &  \\
Sad & 4.205 & 5.205 & -0.192 & -2.819 & 0.007 & ** \\ \bottomrule
\end{tabular}%
}
\caption{User Study Results - Simulation-only. \textbf{Situation 2.} Improvement after the training for Treatment and Control groups. Significance means there is a significant increase of self-reported efficacy or emotions after training.}
\label{tab:app-user-study-TC-survey-S2}
\end{table}

\begin{table}[h!]
\resizebox{0.8\textwidth}{!}{%
\begin{tabular}{@{}lllllll@{}}
\toprule
emotion & treatment\_diff & control\_diff & effect\_size & T-C\_t\_stat & T-C\_p\_value & symbol \\ \midrule
Confident & 1.71 & 0.70 & 0.57 & 2.652 & 0.010 & ** \\
Worried & -1.64 & -1.34 & -0.15 & -0.707 & 0.482 &  \\
Hopeful & 0.55 & 0.36 & 0.10 & 0.463 & 0.644 &  \\
Motivated & 1.05 & 0.36 & 0.36 & 1.683 & 0.096 &  \\
Anger & -0.90 & -0.59 & -0.15 & -0.710 & 0.479 &  \\
Fear & -2.02 & -1.16 & -0.51 & -2.357 & 0.021 & * \\
Disgust & -0.67 & -0.45 & -0.13 & -0.586 & 0.559 &  \\
Sad & -1.43 & -0.89 & -0.30 & -1.400 & 0.165 &  \\ \bottomrule
\end{tabular}%
}
\caption{User Study Results. Difference in difference for Situation 1. Significant result means treatment group and control group are significantly different.}
\label{tab:app-user-study-did-S1}
\end{table}

\begin{table}[h!]
\resizebox{0.8\textwidth}{!}{%
\begin{tabular}{@{}lllllll@{}}
\toprule
emotion & treatment\_diff & control\_diff & effect\_size & T-C\_t\_stat & T-C\_p\_value & symbol \\ \midrule
Confident & 0.71 & 1.18 & -0.24 & -0.686 & 0.498 &  \\
Worried & -1.65 & -0.65 & -0.50 & -1.471 & 0.151 &  \\
Hopeful & 0.53 & 0.41 & 0.05 & 0.159 & 0.874 &  \\
Motivated & -0.18 & 0.47 & -0.37 & -1.074 & 0.291 &  \\
Anger & -0.12 & -0.88 & 0.39 & 1.135 & 0.265 &  \\
Fear & -1.53 & -0.82 & -0.31 & -0.902 & 0.374 &  \\
Disgust & -0.35 & -0.41 & 0.03 & 0.101 & 0.920 &  \\
Sad & -0.59 & -1.29 & 0.38 & 1.112 & 0.275 &  \\ \bottomrule
\end{tabular}%
}
\caption{User Study Results. Difference in difference for Situation 2. Significant result means treatment group and control group are significantly different.}
\label{tab:app-user-study-did-S2}
\end{table}

\section{System prompts used in \ourmodel}
\label{app:system_prompts}
\footnotesize
\begin{tabular}{|p{0.5in}|p{5.9in}|}
\toprule
Skill     & System Prompt \\ \midrule
Describe  & You will be given a context and a utterance, from a conversation that happened in the given context. Does the given utterance describe the given context? To be considered "describe", the utterance needs to stick to the facts, make no judgmental statements, and be objective. Rating Rubric: A "Strong Describe" rating indicate that the utterance is or contains a description of the given context. It sticks to the facts, makes no judgemental statements, and is objective. Do ALL of the following three steps. Step 1: Generate "Reasoning for rating". Step 2: Generate "Describe Rating" in "Strong Describe", "Weak Describe" or "No Describe". A "Weak Describe" rating indicates that the utterance is or contains a description of the given context, but needs improvement since it may not stick to the fact, makes some judgemental statements, or is not fully objective. A "No Describe" rating indicates that the utterance does not describe any aspect of the given context at all. Step 3: Provide additional comments on the ratings similar to the examples given. Finish each step with \#\#\#. Twenty words minimum. YOU MUST FINISH EACH STEP WITH \#\#\#  \\ \midrule
Express   & You will be given a context and a utterance, from a conversation that happened in the give context. Does the given utterance explicitly express how the speaker feel in the conversation? To be considered "express", the utterance needs to EXPLICITLY express your feelings about the given context, including things like "this makes me feel", "I feel ... by your actions", "this situation/your action... has caused me..." with adjectives or nouns describing emotions. Do ALL of the following three steps. Step 1: Generate "Reasoning for rating". Step 2: Generate "Express Rating" in "Strong Express", "Weak Express" or "No Express". Rating Rubric: A "Strong Express" rating indicate that the utterance is or contains a EXPLICIT expression of the felt emotions. YOU CANNOT INTERPRET THE SENTIMENT IF THE SPEAKER DOES NOT MENTION EMOTIONS. A "Weak Express" rating indicates that the utterance is or contains an expression of your feelings or opinions about the given context, but can be made more explicit in expressing feelings. A "No Express" rating indicates that the utterance does not express your feelings or opinions about the given context at all. Step 3: Provide additional comments on the ratings similar to the examples given. Finish each step with \#\#\#. Twenty words minimum. YOU MUST FINISH EACH STEP WITH \#\#\# \\ \midrule
Assert    & You will be given a context and a utterance, from a conversation that happened in the give context. Does the given utterance assert your needs or wants about the given context? To be considered "assert", the utterance needs to be asking for what you want or saying no clearly. Do ALL of the following three steps. Step 1: Generate "Reasoning for rating". Step 2: Generate "Assert Rating" in "Strong Assert", "Weak Assert" or "No Assert". Rating Rubric: A "Strong Assert" rating indicate that the utterance is or contains an assertion of your needs or wants about the given context. A "Weak Assert" rating indicates that the utterance is or contains an assertion of your needs or wants about the given context, but needs improvement in making it more explicit or stronger. A "No Assert" rating indicates that the utterance does not contain an assertion of the needs or wants. Step 3: Provide additional comments on the ratings similar to the examples given. Finish each step with \#\#\#. Twenty words minimum. YOU MUST FINISH EACH STEP WITH \#\#\#  \\ \midrule
Reinforce & You will be given a context and a utterance, from a conversation that happened in the give context. Does the given utterance reinforce your needs or wants about the given context? To be considered "reinforce", the utterance needs to reinforce some reward for the other person. Do ALL of the following three steps. Step 1: Generate "Reasoning for rating". Step 2: Generate "Reinforce Rating" in "Strong Reinforce", "Weak Reinforce" or "No Reinforce". Rating Rubric: A "Strong Reinforce" rating indicate that the utterance is or contains a reinforcement for the other person about the given context. A "Weak Reinforce" rating indicates that the utterance is or contains a reinforcement of your needs or wants about the given context, but needs improvement, for example, it may not be a reward for the other person or it is not communicated clearly. A "No Reinforce" rating indicates that the utterance does not have a reinforcer for the other person. Step 3: Provide additional comments on the ratings similar to the examples given. Finish each step with \#\#\#. Twenty words minimum. YOU MUST FINISH EACH STEP WITH \#\#\#   \\\midrule

Mindful   & You will be given a context and a utterance, from a conversation that happened in the give context. Does the given utterance show the speaker is being mindful? To be considered "mindful", the utterance needs to be stick to the speaker\textbackslash{}'s goal and does not get distracted by what the other person says. Do ALL of the following three steps. Step 1: Generate "Reasoning for rating". Step 2: Generate "Mindful Rating" in "Yes" or "No". Rating Rubric: A "Yes" rating indicate that the utterance is showing mindfulness. A "No" rating indicates that the utterance shows a lack of mindfulness, the speaker may be responding to attacks or losing track of their goals. Step 3: Provide additional comments on the ratings similar to the examples given. Finish each step with \#\#\#. Twenty words minimum. YOU MUST FINISH EACH STEP WITH \#\#\#  \\\midrule
Confident & You will be given a context and a utterance, from a conversation that happened in the give context. Does the given utterance show the speaker is being confident? To be considered "confident", the utterance needs to have a confident tone, is effective and competent in conveying the speaker\textbackslash{}'s goal. Do ALL of the following three steps. Step 1: Generate "Reasoning for rating". Step 2: Generate "Confident Rating" in "Yes" or "No". Rating Rubric: A "Yes" rating indicate that the utterance is showing confidence. A "No" rating indicates that the utterance shows a lack of confidence. Step 3: Provide additional comments on the ratings similar to the examples given. Finish each step with \#\#\#. Twenty words minimum. YOU MUST FINISH EACH STEP WITH \#\#\#  \\\midrule
Negotiate & You will be given a context and a utterance, from a conversation that happened in the give context. Does the given utterance contain a negotiation? To be considered "negotiate", the utterance needs to offer and ask for other solutions in the given context. Do ALL of the following three steps. 1) Generate "Reasoning for rating", 2) Generate "Negotiate Rating" in "Strong Negotiate", "Weak Negotiate" or "No Negotiate". Rating Rubric: A "Strong Negotiate" rating indicate that the utterance offers or asks clearly for an alternative solution. A "Weak Negotiate" rating indicates that the utterance is or contains a negotiation of your needs or wants about the given context, but may not be clear enough and needs improvement. A "No Negotiate" rating indicates that the utterance does not contain any negotiation at all. 3) Provide additional comments on the ratings similar to the examples given. Finish each step with \#\#\#. Twenty words minimum. YOU MUST FINISH EACH STEP WITH \#\#\#     \\ \bottomrule
\end{tabular}
\input{100_App_03data}
\newpage
\section{User Study Interface}
\label{app:user-study-interface}
\subsection{Simulation+Feedback group, Training Conversation - part 1}
\begin{figure}[htbp]
    \centering
    \includegraphics[width=.93\textwidth]{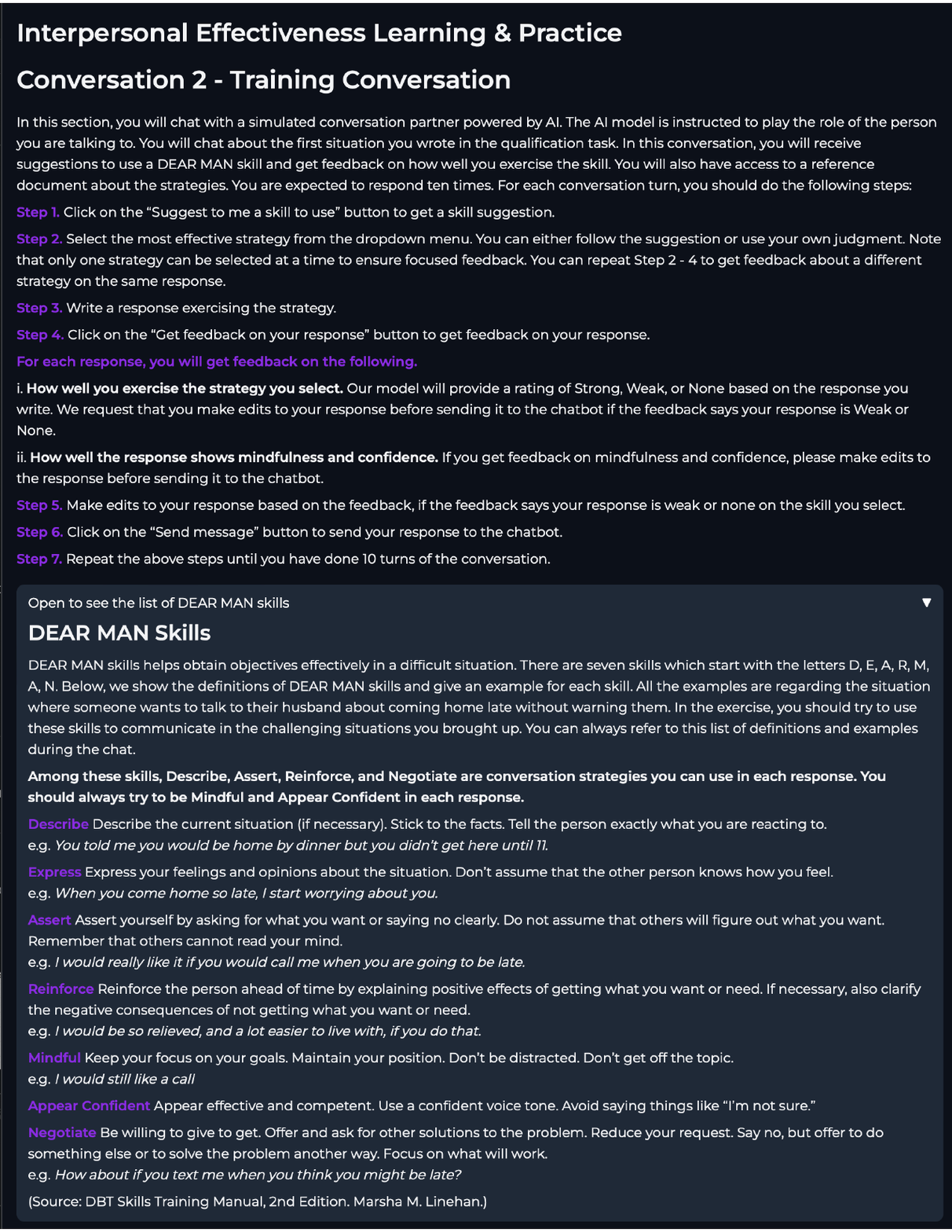}
\end{figure}
\newpage
\subsection{Simulation+Feedback group, Training Conversation}
\begin{figure}[h]
    \centering
    \includegraphics[trim={1mm 0 1mm 0},clip,height=.78\textheight]{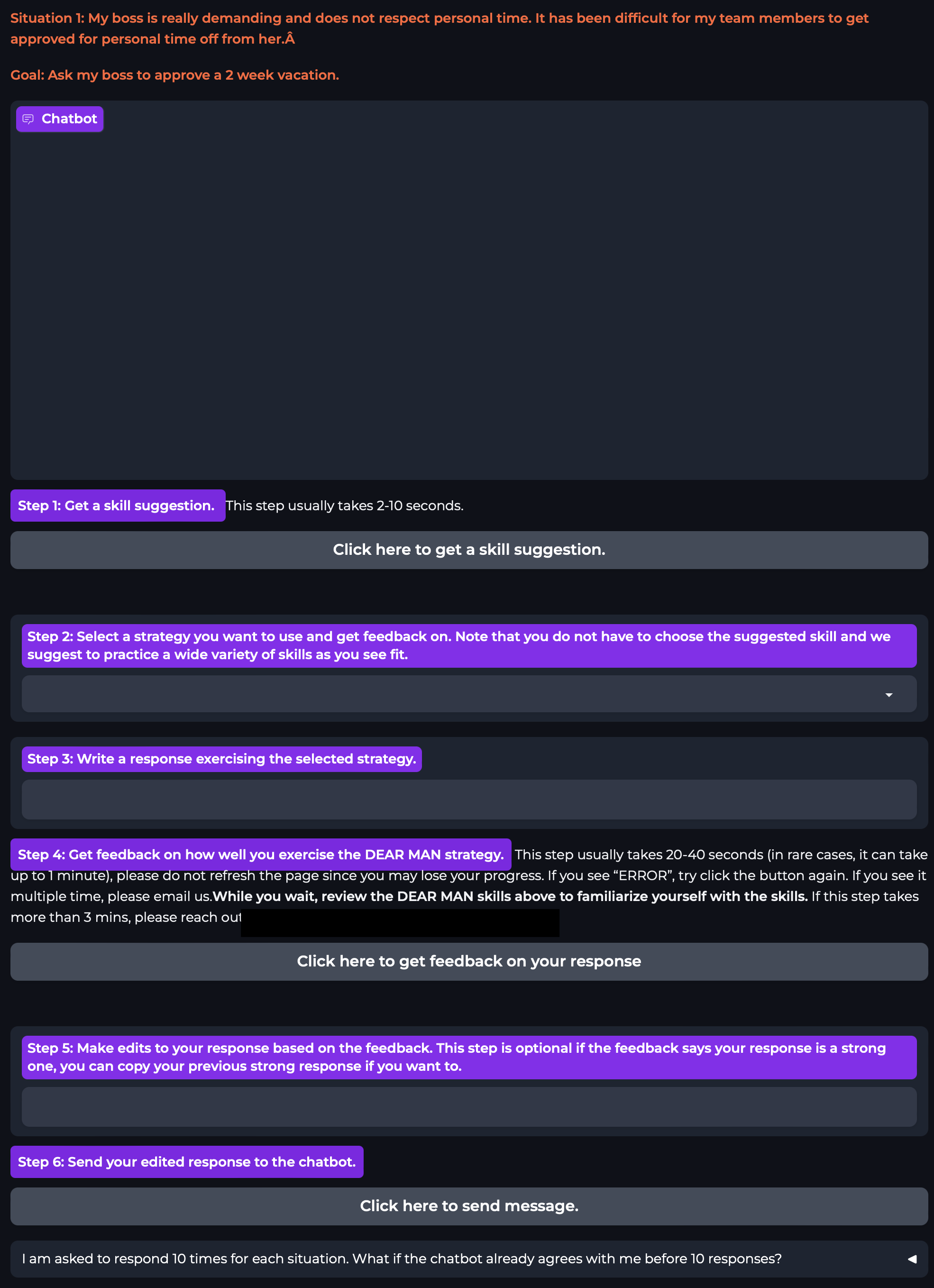}

\end{figure}

\newpage
\subsection{Simulation-only group, Training Conversation part 2}
\begin{figure}[h]
    \centering
    \includegraphics[trim={26mm 0 26mm 0},clip,height=.77\textheight]{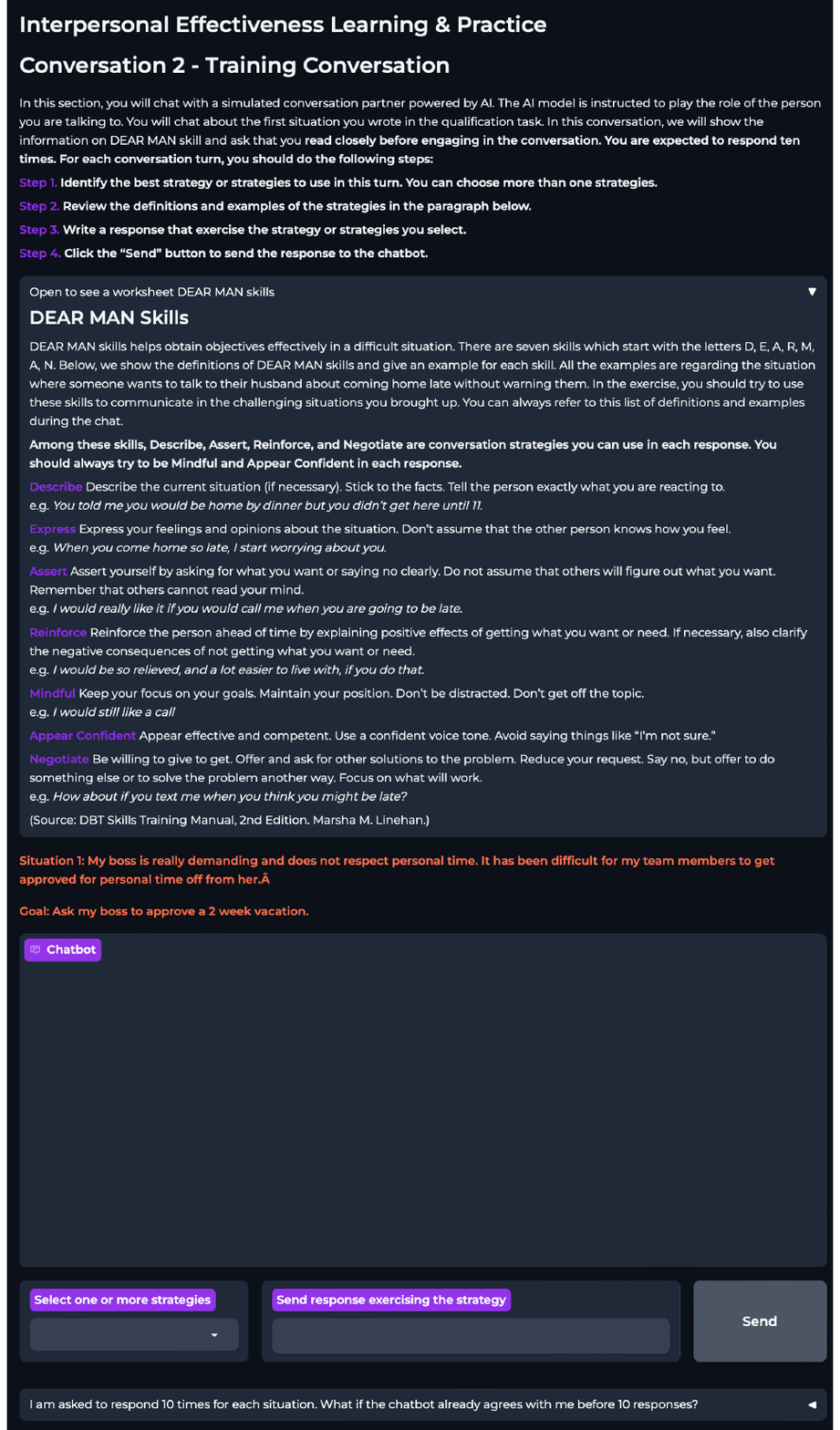}

\end{figure}

\newpage
\subsection{Evaluation Conversation}
\begin{figure}[h]
    \centering
    \includegraphics[trim={10mm 0 10mm 0},clip,height=.78\textheight]{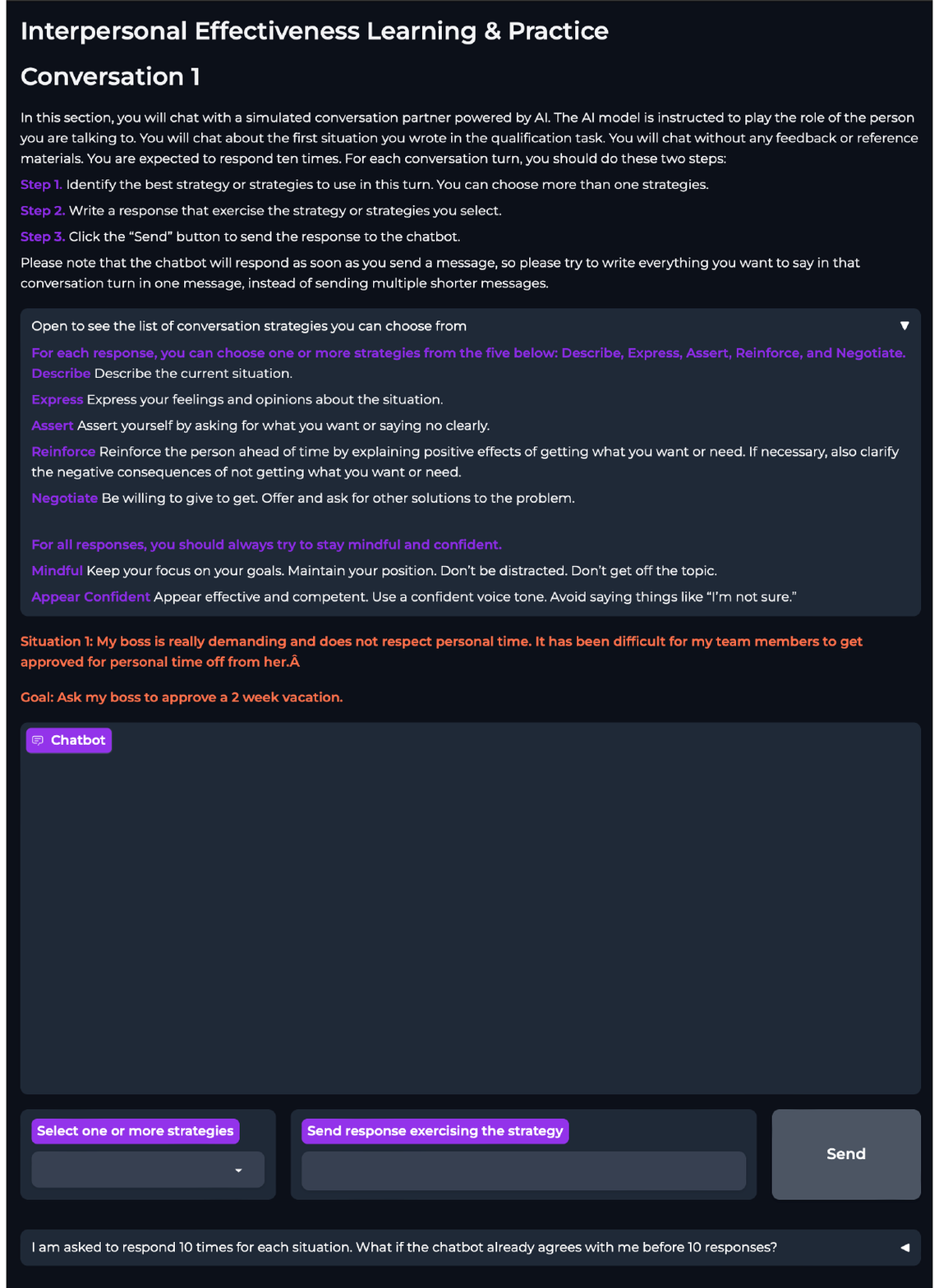}
\end{figure}

\clearpage
\section{User Study Results - Skill Mastery by Skill}
\begin{figure}[h]
    \centering
    \includegraphics[width=\columnwidth]{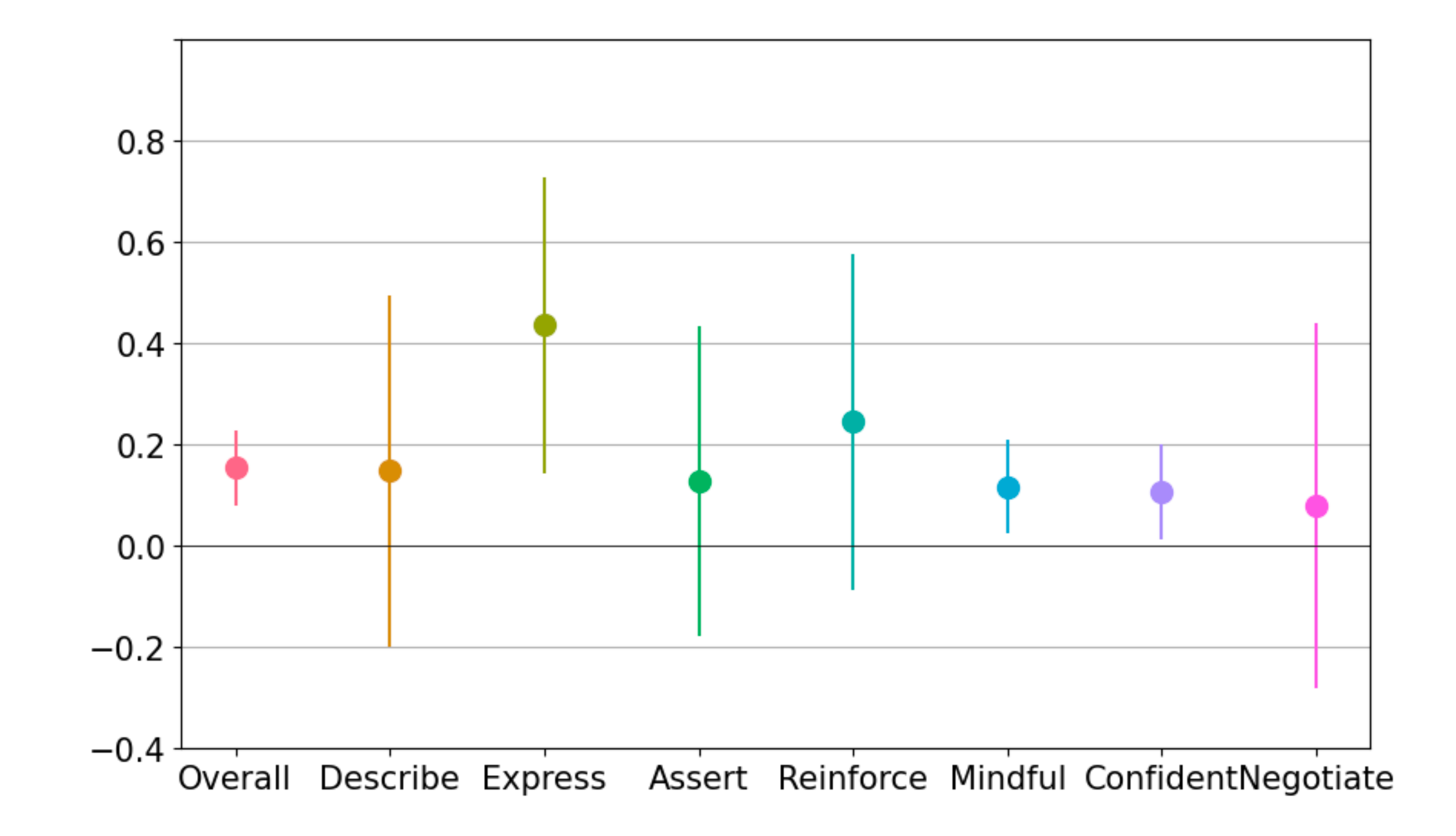}
    \caption{Difference between Simulation+Feedback group and Simulation-only group on the improvement of skill use by each skill. We use bootstrapping to estimate confidence intervals (5000 iterations). Simulation+Feedback group sees a significantly higher increase in overall skill use (15.6\%, $p=.000$), Express(43.2\%, $p=.003$), Mindful(11.6\%, $p=.012$), and Confident skills(10.8\%, $p=.021$). ***: $p<.001$, **:$p<.01$, *:$p<.05$, $d$: Cohen's $d$. }
    \label{app:utterance-level-comparison}
\end{figure}
\clearpage

\section{Expert Data Annotation Interface}
\label{appendix:sec:expert-interface}
\begin{figure}[h]
    \includegraphics[width=0.8\textwidth]{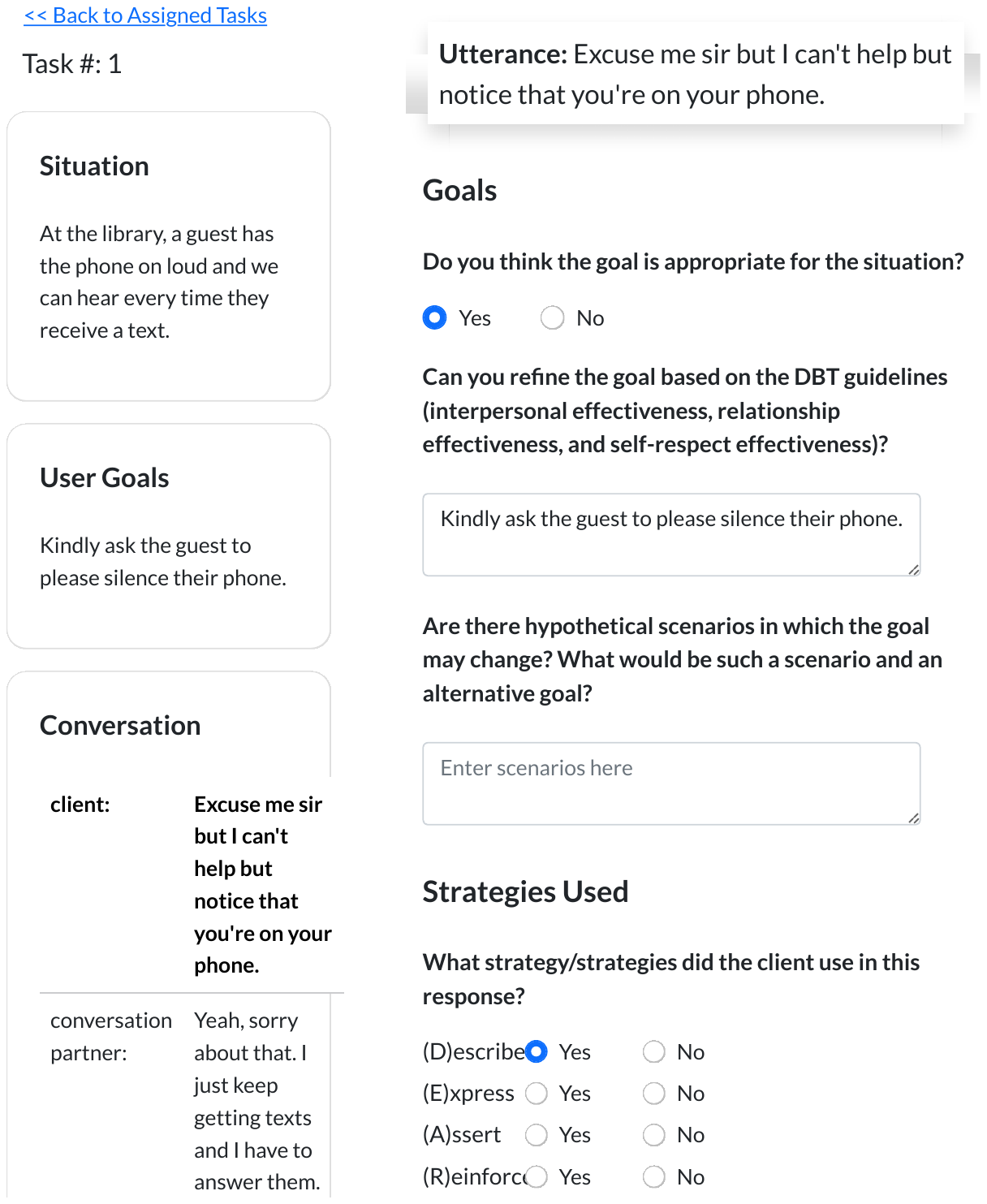}
    \caption{Screenshot of the interface used for expert data annotation. Continues on the next page (1/4). }
\end{figure}

\begin{figure}[htbp]
    \includegraphics[width=\textwidth]{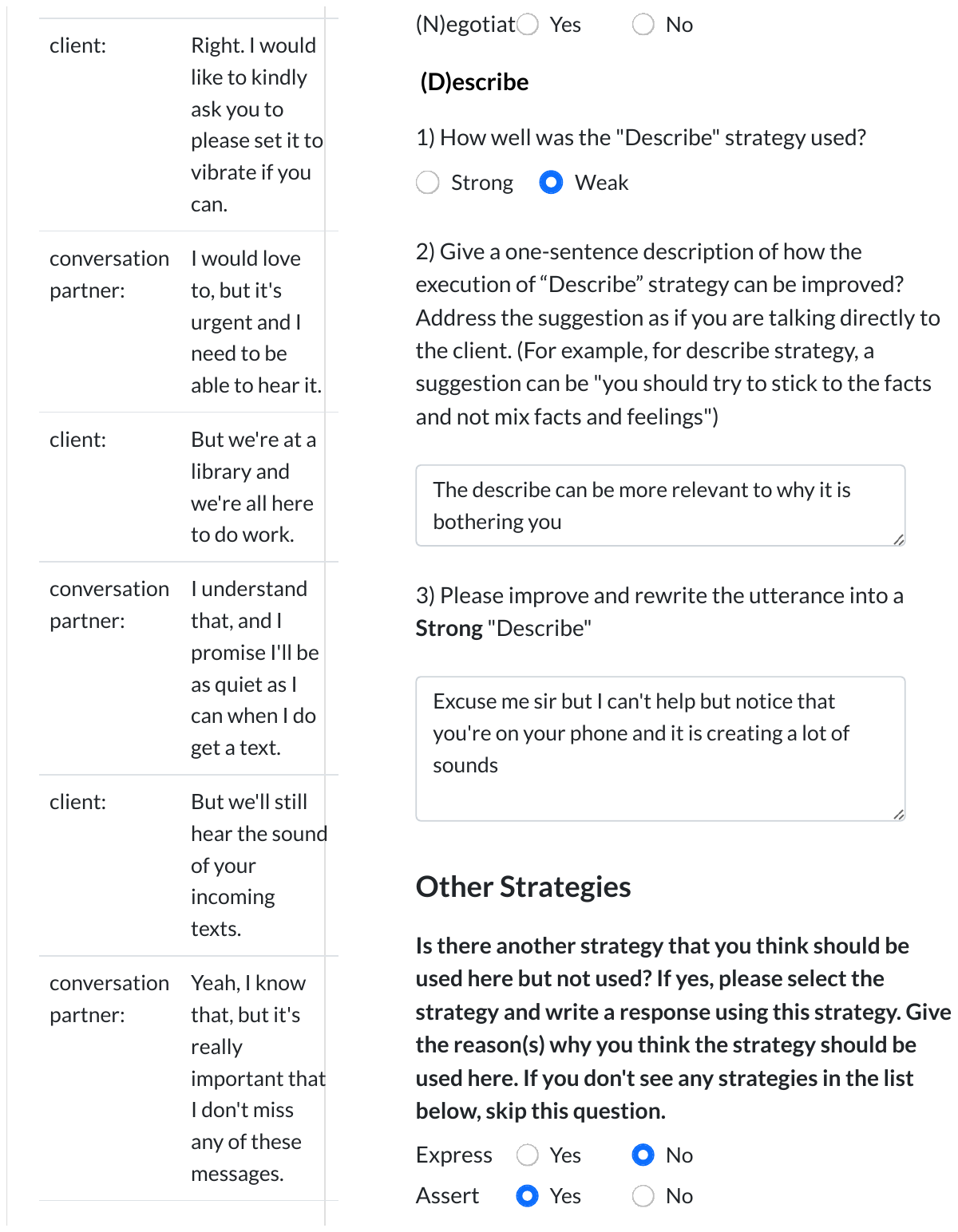}
    \caption{Screenshot of the interface used for expert data annotation. Continues on the next page (2/4). }
\end{figure}

\begin{figure}[htbp]
    \includegraphics[width=\textwidth]{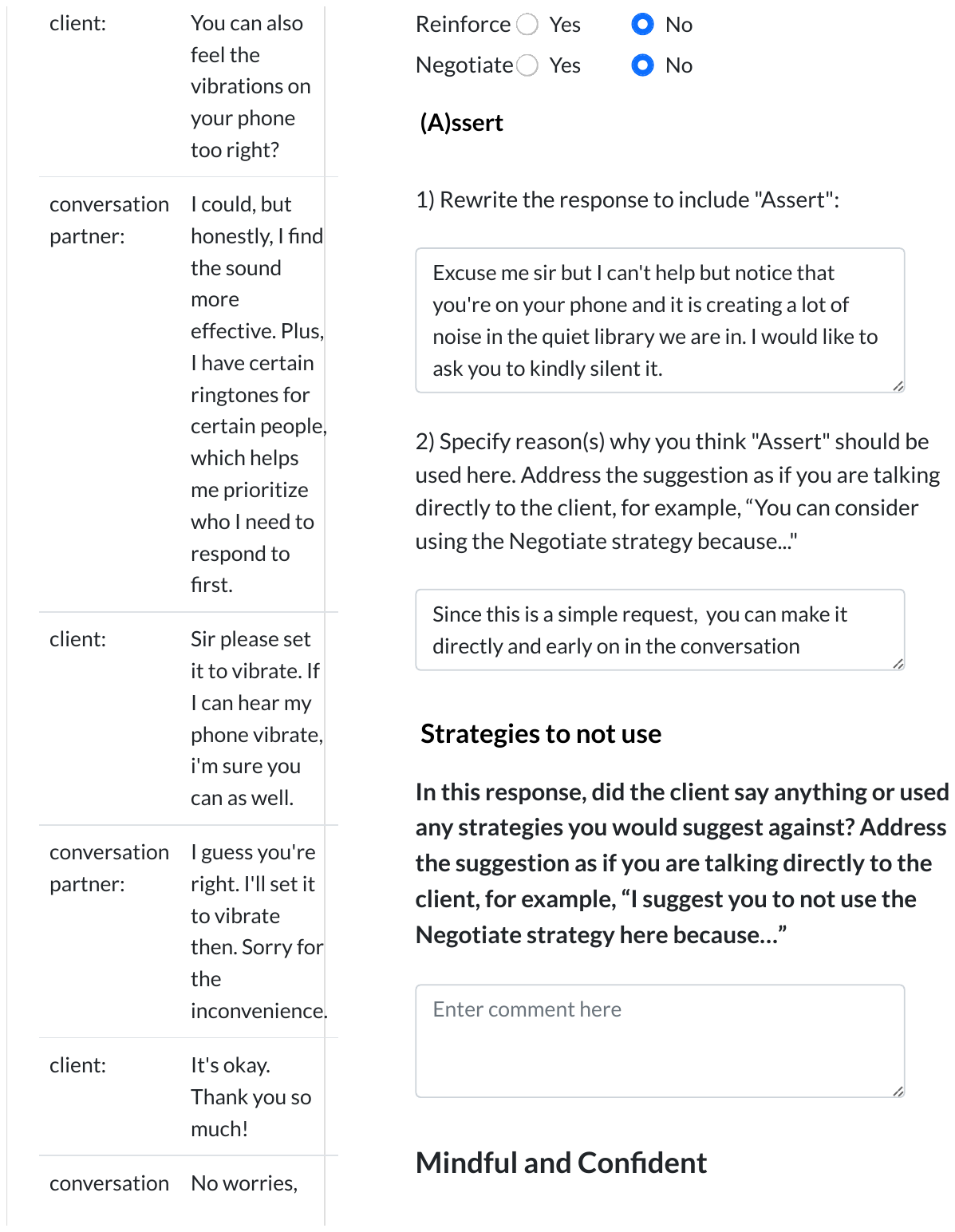}
    \caption{Screenshot of the interface used for expert data annotation. Continues on the next page (3/4). }
\end{figure}

\begin{figure}[htbp]
    \includegraphics[width=\textwidth]{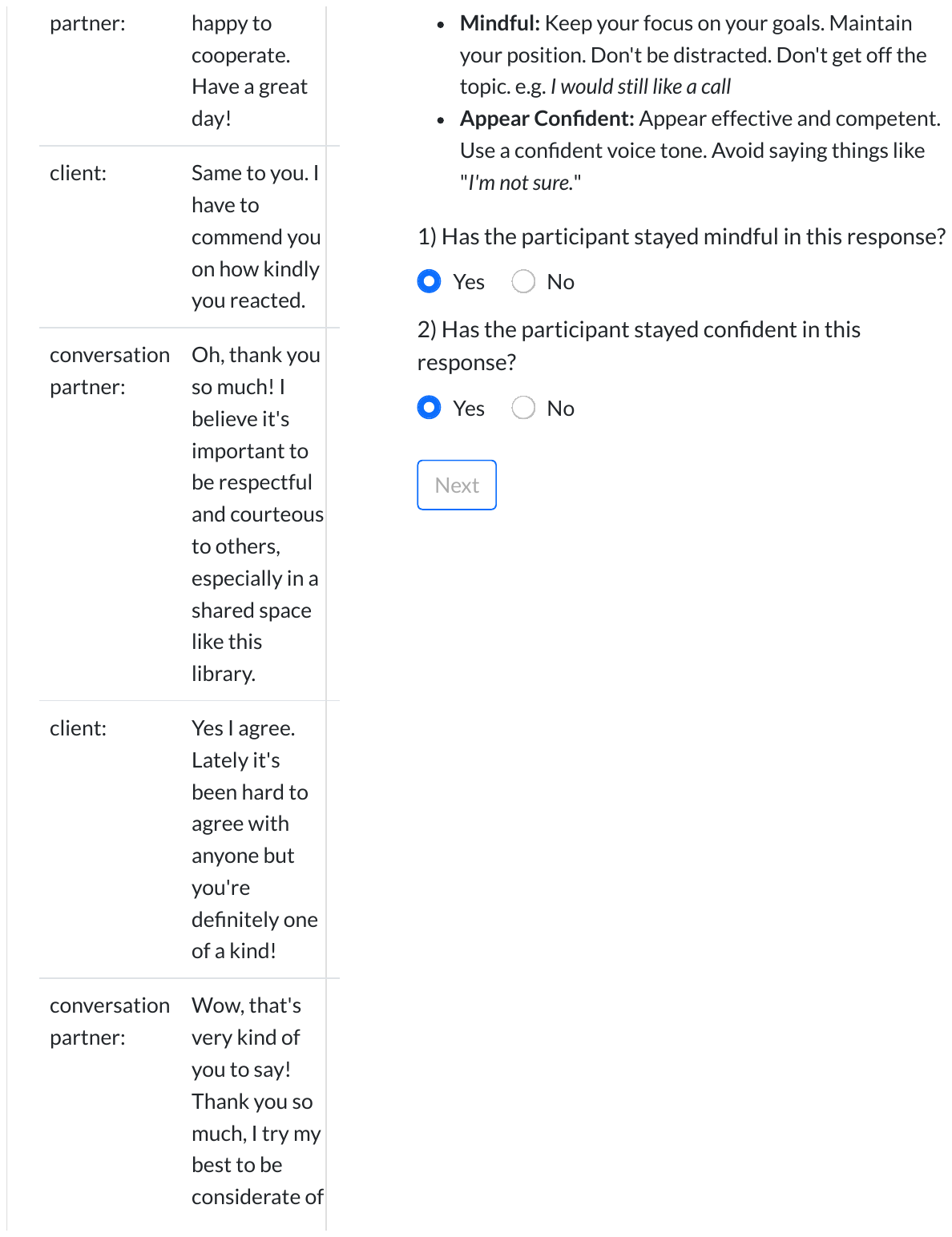}
    \caption{Screenshot of the interface used for expert data annotation (4/4). }
\end{figure}

%% file: 100_App_02formative.tex
\section{DEAR MAN Definition~\cite{linehan2014dbt}}
\label{app:dearman-worksheet}
\begin{table}[h]
\resizebox{\textwidth}{!}{%
\begin{tabular}{l|l}
\hline
\textbf{DEAR MAN Skill} & \textbf{Definition and Examples} \\ \hline
Describe & \begin{tabular}[c]{@{}l@{}}Describe the current situation (if necessary). Stick to the facts. Tell the person exactly what you are reacting to. \\ e.g. You told me you would be home by dinner but you didn't get here until 11.\end{tabular} \\ \hline
Express & \begin{tabular}[c]{@{}l@{}}Express your feelings and opinions about the situation. Don't assume that the other person knows how you feel. \\ e.g. When you come home so late, I start worrying about you.\end{tabular} \\ \hline
Assert & \begin{tabular}[c]{@{}l@{}}Assert yourself by asking for what you want or saying no clearly. Do not assume that others will figure out \\what you want. Remember that others cannot read your mind. \\ e.g. I would really like it if you would call me when you are going to be late.\end{tabular} \\ \hline
Reinforce & \begin{tabular}[c]{@{}l@{}}Reinforce the person ahead of time by explaining positive effects of getting what you want or need. If necessary, \\also clarify the negative consequences of not getting what you want or need. \\ e.g. I would be so relieved, and a lot easier to live with, if you do that.\end{tabular} \\ \hline
Mindful & \begin{tabular}[c]{@{}l@{}}Keep your focus on your goals. Maintain your position. Don't be distracted. Don't get off the topic.\\ e.g. I would still like a call\end{tabular} \\ \hline
Appear Confident & Appear effective and competent. Use a confident voice tone. Avoid saying things like "I'm not sure." \\ \hline
Negotiate & \begin{tabular}[c]{@{}l@{}}Be willing to give to get. Offer and ask for other solutions to the problem. Reduce your request. Say no, but offer \\to do something else or to solve the problem another way. Focus on what will work. \\ e.g. How about if you text me when you think you might be late?\end{tabular} \\ \hline
\end{tabular}
}

\end{table}

\section{Formative Study Details}
\label{app:formative_study}

We recruit from the clinical psychology departments in four universities and select who indicated in the sign up form that they ``sometimes'' or ``regularly'' work with clients on DEAR MAN skills(~\S\ref{app:expert-recruitment-question}). We conduct the study with three clinical experts (E1, E2, E3) in semi-structured interviews~\cite{kallio2016systematic,Yin2020RoleOT}. We first show the experts a preliminary version of the interface, of which the main functions include: collecting information from users about a difficult situation they face and a LM-backed chatbot that is instructed to roleplay the conversation partner in the user-specified situation who is difficult to convince. We first let the expert try the interface, followed by structured questions on skill teaching, learning, and practice, and the measurement of success in DEAR MAN skill acquisition. We share the insights that informed us about several design decisions. Clinical experts are paid \$37.5/hour for this two hour task.

\section{Expert Recruitment - DEAR MAN experience question}
\label{app:expert-recruitment-question}
In the expert signup form, we specifically ask for their experience with DBT and DEAR MAN skills. We only selected those who chose ``5 - I sometimes work with clients on DEAR MAN in my practice'' or ``6 - I regularly work with clients on DERA MAN in my practice''.

\begin{itemize}
    \item 1 - I have only heard about it
    \item 2 - I have learned about it in school / read about it extensively but never used it in practice
    \item 3 - I have worked with clients on DBT but not DEAR MAN specifically
    \item 4 - I have worked with clients on DEAR MAN at least once
    \item 5 - I sometimes work with clients on DEAR MAN in my practice
    \item 6 - I regularly work with clients on DERA MAN in my practice
\end{itemize}

\section{Simulation - System Prompt}
\label{app:system_prompt_simulation}

We use the below prompt as an input to an LM, to generate system prompt for the simulation LM. 

``Situation: My husband always comes home late and he doesn't text me or call me. Prompt: Act like my husband who always comes home late without calling or texting me. Prompt: Act like my boss who regularly calls me on weekends but I don't want to work on the weekends. Situation: My friend has depression and she relies on me 24/7 and I feel drained. Prompt: Act like my friend who has depression and who relies on me whenever you have an issue and I want to convince you to seek professional help and not rely on a friend for all your issues. Situation: My neighbor frequently plays loud music at a late hour and hosts big parties, which affect my sleep. Prompt: Act like my neighbor. You frequently play loud music at a late hour and host big parties. Situation: The airline lost my luggage and the customer service agents have been passing the buck.  Prompt: Act like a customer service agent. Your airline lost my luggage and your colleagues have been passing the buck.''

(We use $\backslash$n in the prompt separating each situation.)

\section{Human Evaluation}
\label{app:human_eval}

We conduct human evaluation with 13 CS PhD students who are familiar with LM generations. The task is to identify if the model-generated feedback is the same with expert-generated feedback. The participants are guided with the rubric below:

\xhdr{1)} Think about if you receive the feedback, would you have acted the same, or differently?\\
\xhdr{2)} Do they give the same \textit{reason} of why the utterance should be improved?\\
\xhdr{3)} Do not think about whether they are equally specific or actionable. (We account for these two measures separately.)\\
\xhdr{4)} If one of the feedback seems to be noisy data, like if they just output a single word ``Describe'', they are not considered the same.\\

The participants discussed a few examples, like the following, at the beginning to help guide the annotation:\\

\textbf{Example 1}\\
\textbf{Expert:} Try using more non-judgmental language here and focus on describing the details - think about what a video camera would see. \\
\textbf{Model:} To improve the utterance should focus on describing the actions without assigning motive or making judgmental statements For example "You expressed dissatisfaction with the gift from your in-laws in public" would be more objective\\
\textbf{Label:} Same. 

\textbf{Example 2}\\
\textbf{Expert:} Staying mindful means holding your ground and avoiding being distracted by the server's comments about how good the other beer tastes.\\
\textbf{Model:} To align with the DEAR MAN framework the utterance could be structured to Describe the situation Express feelings Assert needs Reinforce the positive stay Mindful Appear confident and Negotiate if necessary For example\\
\textbf{Label:} Not the same

The participants are unaware of which source model their data originates from, and none of them have seen any generations from any of the model variants prior to evaluation. Each pair of model-generated and expert feedback receives three annotations. We consider the majority of these three annotations as the final annotation. The average Fleiss' Kappa of the annotations is: 0.667, showing substantial agreement between the annotators.

\section{Specificity and Actionability - Few shot learning prompts}

\textbf{Actionability}

On a scale from 1 to 5, how actionable is the following feedback?

Feedback: Stay on topic and approach the situation with the intention of finding a resolution Consider expressing your concerns and the impact of your mother's actions more calmly and objectively rather than accusing her of negative intentions
Actionability: 4

Feedback: To align with the DEAR MAN framework the utterance could be more assertive and clear about the need and the reason behind the request For example
Actionability: 3

Feedback: I would suggest to use express more heavily to express appreciation and understanding of the other person's point of view.
Actionability: 3

Feedback: To align the utterance with the DEAR MAN framework which is a skill from Dialectical Behavior Therapy (DBT) used to teach effective communication the speaker could structure their statement with more clarity and respect focusing on the following components
Actionability: 2

Feedback: You can try phrasing your ask more assertively, using "I want" rather than "should".
Actionability: 5

Feedback: 1 **Describe**
Actionability: 1

Feedback: To align the utterance more closely with the DEAR MAN framework which is a skill from Dialectical Behavior Therapy (DBT) designed to help people communicate effectively and assertively the utterance could be structured as follows
Actionability: 1

\textbf{Specificity}

On a scale from 1 to 5, how specific is the following feedback, given the situation and the utterance?

Situation: My colleague keep borrowing money from me without completely paying her old debts, and she don't feel ashamed to come asking despite I've confronted her several times about it. But I don't know her to refuse lending to her because I have it and can't lie I don't have any money 
Utterance: I'm not happy with you any longer because you're owing me a lot of money promising to pay back soon but you haven't.
Feedback: Be more specific with how much money, and how often, your friend is borrowing.
Specificity: 5

Situation: my friend borrowed my car and hid the fact that he had a slight accident and dent my car
Utterance: oh yes you did and you know it, i already found out the truth because Jenny told me what you guys did the other day, but here you are lying to my face
Feedback:  This utterance demonstrates mindfulness by focusing on the issue of the car accident and the dishonesty rather than getting sidetracked by other topics It's direct and addresses the core issue effectively
Specificity: 4

Situation: I went to dinner with my friends and a waiter brought me the wrong beer for the second time. I had asked for a Blue Moon but they kept bringing me Samuel Adams.
Utterance: No worries. Why no Blue Moon? I'm just curious.
Feedback: The speaker maintains composure and expresses curiosity rather than frustration or anger indicating mindfulness in addressing the mistake without getting sidetracked by emotions
Specificity: 3

Situation: At the library, a guest has the phone on loud and we can hear every time they receive a text.
Utterance: But we'll still hear the sound of your incoming texts.
Feedback: This utterance is appropriate as it is It objectively describes the situation without adding any unnecessary judgment or emotion
Specificity: 2

Situation: My colleague keep borrowing money from me without completely paying her old debts, and she don't feel ashamed to come asking despite I've confronted her several times about it. But I don't know her to refuse lending to her because I have it and can't lie I don't have any money
Utterance: I'm not happy with you any longer because you're owing me a lot of money promising to pay back soon but you haven't.
Feedback: To align the utterance more closely with the DEAR MAN framework which is a skill from Dialectical Behavior Therapy (DBT) designed to help people communicate effectively and assertively the utterance could be structured as follows
Specificity: 1

%% file: 100_App_03data.tex
\section{mTurk and Prolific Recruitment}
\label{app:mturk}
Participants were paid \$15/hour for both data collection and randomized trial study. To incentivize skill learning, we also pay an additional \$10 bonus to the top 30\% in each S and S+F groups, who exhibit the highest levels of skill use, rated by our model.

\subsection{Qualification Task Posting}
Thank you for clicking into this qualification task! We are looking for people to chat with our chatbot as part of our data collection. 

In the actual task, you will be asked to complete three chats (10 responses each), for about 45 minutes (We will be paying about \$15/hour!). We will ask you to describe three situations where you find difficult to communicate in and the chatbot will simulate the person you will be talking to (no personal information will be collected). You will be asked to select communication strategies you used in each response, like "describe situation", "express feelings", "negotiate", etc.

If you are interested in the actual task, please complete this qualification HIT! Here, you will be asked to describe one situation. You will be able to re-use the answer here in the actual task.

\textbf{With 1-2 sentences, describe a situation that you find difficult to communicate in.}

- Please clearly state the nature of your connection with the person you are communicating, such as "my husband" or "my boss", while avoiding disclosing any identifiable personal information, such as names, locations, etc. 

- Please provide information regarding the factors contributing to the challenging situation, such as past instances of unsuccessful communication or anticipated behaviors.

Example 1: My husband always comes home late without giving me a notice and despite my effort to talk to him, he does not change.

Example 2: My boss is really demanding and does not respect personal time. It has been difficult for my team members to get approved for personal time off from her. 

\textbf{What is the goal of the conversation? }

Example 1: Convince my husband to call me next time when he needs to come home late.

Example 2: Ask my boss for approval of a week long vacation. 

\subsection{Data Collection Task Posting}

\textbf{Study Description}

In this study, you will complete three tasks. In each task, you will describe a difficult situation in one of social, work, and family categories where you find it difficult to communicate with someone. (Similar to what you did in the qualification task, and you can re-use the examples you gave in the qualification task.) Then, you will chat with a chatbot (powered by AI) who will play the role of the conversation partner and you will try to achieve your conversation goal. You will be asked to respond 10 times, or until the chatbot agrees with you, whichever comes first. In each response, we ask you to select communication strategies that you used in that message. More details will be given in the link.

Please note that the chatbot will respond as soon as you send a message, so please try to write everything you want to say in that conversation turn in one message, instead of sending multiple shorter messages.
 
\textbf{IMPORTANT: How do I confirm the completion of this task?\\}
For each task, you will be provided a TASK CODE (6 letters) and you will be asked to copy paste the TASK CODE in the boxes below.\\
Please try to finish each situation in one go (expect it to be around 10-15 minutes for each situation). If you exit, you may lose the TASK CODE and may have to start from the beginning.\\
Please note that you will only get the payment if you complete the entire study, i.e. 10 responses or until the chatbot agrees with you for all three situations.\\
If you experience any technical difficulties, please reach out to xxx@xxx.com
 
\textbf{Task Instruction and Example\\}
When you open each task link, you will see step-by-step instructions and examples. The same information can be accessed at: xxx@xxx.com\\
Provide the TASK 1 CODE here:\\
Provide the TASK 2 CODE here:\\
Provide the TASK 3 CODE here:

\subsection{User Study Qualification}
\textbf{Introduction}

Thank you for clicking into this qualification task! We are looking for people who want to improve their communication skills by chatting with our chatbot . 

Have you ever had a difficult conversation with someone or avoided having a conversation with someone because you were afraid that it might not go well? We’re designing a tool that can help people to confidently communicate with others in these difficult situations. In the main task, you will be asked to complete 4 chats (10 responses each), for about an hour. We will be paying about \$15/hour with \$10 bonus for top 30\%! We will ask you to describe two situations in which you find it difficult to communicate, and the chatbot will simulate the person you will be talking to (no personal information will be collected in the chats). The material in this qualification task will be automatically loaded into the main task.

If you are interested in the main task, please complete this qualification task! Here, you will be asked to describe two situations, communication goals, and rate how difficult you think they are.

You will be qualified as long as the situations, goals and difficulty levels are reasonable. 

If you are interested in the main task and are not able to complete this task due to mTurk qualifications, please email xxx@xxx.com with your answers. We will give full consideration to answers received via email.

\textbf{Task description}
With 1-2 sentences, describe two situations that you find difficult to communicate in. You should consider both situations to be difficult, situations that are too easy will not be accepted.\\
\textbf{Requirements:}\\
You must clearly state the nature of your connection to the person with whom you are communicating, such as “my husband” or “my boss”, while avoiding disclosing any identifiable personal information, such as names, locations, etc. \\
You should provide as much information as possible regarding the factors contributing to the challenging situation, such as past instances of unsuccessful communication or anticipated resistance behaviors.\\
Example 1: \\
Situation: My husband always comes home late without giving me a warning and despite my effort to talk to him, he does not change.\\
Goal: Convince my husband to let me know in advance when he needs to arrive late.\\
Example 2: \\
Situation: My boss is really demanding and does not respect personal time. It has been difficult for my team members to get approved for personal time off from her. \\
Goal: Get approval from my boss for a two-week vacation while maintaining a positive and professional relationship.\\
\textbf{Your answers here}\\
\textbf{Situation 1:} \\
 \textbf{Goal 1:}\\
 \textbf{On a scale of 1-9, how difficult is it for you to communicate in this situation? Note we require both situations to be at least 7 - Difficult.}\\
1 - Extremely Easy\\
2 - Very easy\\
3 - Easy\\
4 - Somewhat easy\\
5 - Neither Difficult nor Easy\\
6 - Somewhat difficult\\
7 - Difficult\\
8 - Very difficult\\
9 - Extremely difficult\\
\textbf{Situation 2:} \\
\textbf{ Goal 2}:\\
\textbf{On a scale of 1-9, how difficult is it for you to communicate in this situation? Note we require both situations to be at least 7 - Difficult.}\\
1 - Extremely Easy\\
2 - Very easy\\
3 - Easy\\
4 - Somewhat easy\\
5 - Neither Difficult nor Easy\\
6 - Somewhat difficult\\
7 - Difficult\\
8 - Very difficult\\
9 - Extremely difficult\\
\subsection{Randomized Trial Posting}
\textbf{Study Description}\\
Congratulations on getting selected to participate in this study! We are a group of researchers building a tool to help people improve interpersonal communication skills, with the help of Artificial Intelligence. In this study, you will interact with our chatbots, answer some questions about the situations you wrote in the qualification task (this information will be preloaded into the study website), get detailed feedback on your conversation responses, and learn and improve communication skills! \\

BONUS information:  You will receive a bonus of \$10 if you are at the top 30\% of the participants in terms of how well you exhibit the skills taught in the tool - more information in the study link.\\

\textbf{IMPORTANT: How do I confirm the completion of this task?}\\
For each task, you will be provided a COMPLETION CODE (6 letters) at the end of the study and you will be asked to provide this code in the box below.\\
Please try to finish this study in one go (expect it to be around one hour). If you exit, you may lose the progress and may have to start from the beginning.\\
Please note that you will only get the payment if you complete the entire study. \\
Please note that the link expires in 72 hours so please allocate an hour in the following 72 hours to complete this study. If this time frame does not work for you, I am happy to share an alternative link at your desired time, please email me if that is the case \\
If you experience any technical difficulties, please reach out to xxx@xxx.com\\
Sincerely appreciate your participation!\\
Provide the COMPLETION CODE here:\\
Provide the survey code here:\\

\newpage
\subsection{User Demographics}

\begin{table}[h]
\resizebox{.5\textwidth}{!}{%
\begin{tabular}{ll|ll|ll}
\hline
\multicolumn{2}{l|}{Gender} & \multicolumn{2}{l|}{Age} & \multicolumn{2}{l}{Race/Ethnicity}         \\ \hline
Man                & 63.8\% & 18-24      & 44.7\%     & White                                & 55.3\% \\
Woman              & 36.2\% & 25-34      & 44.7\%     & Black           & 25.5\% \\
  &                         & 35-44      & 8.5\%     & Asian                              & 10.6\%  \\
&        & 45-54      & 2.1\%     & Other   & 4.3\%  \\
&        &     &       & Mixed & 4.3\%  \\
\hline
\end{tabular}
}
\caption{Breakdown of participant demographics by gender, age, and race/ethnicity- Randomized Trial, Prolific.}
  \vspace{-0.2in}
\label{tab:demographics}
\end{table}

\begin{table}[h]
\resizebox{.8\textwidth}{!}{%
\begin{tabular}{ll|ll|ll}
\hline
\multicolumn{2}{l|}{Gender} & \multicolumn{2}{l|}{Age} & \multicolumn{2}{l}{Race/Ethnicity}         \\ \hline
Man                & 46.2\% & 18-24      & 15.4\%     & White                                & 65.4\% \\
Woman              & 50.0\% & 25-34      & 26.9\%     & Hispanic or Latino          & 11.5\% \\
Undiscl.  & 3.8\%  & 35-44      & 57.7\%     & Asian                              & 11.5\%  \\
&        &      &      & American Indian / Alaskan Native   & 7.7\%  \\
&        &    &       & Black or African American (Non-Hispanic) & 3.8\%  \\
\hline
\end{tabular}}
\caption{Breakdown of participant demographics by gender, age, and race/ethnicity - Data Collection.}
  \vspace{-0.2in}
\label{tab:demographicsMturk1}
\end{table}

\begin{table}[h]
\resizebox{.8\textwidth}{!}{%
\begin{tabular}{ll|ll|ll}
\hline
\multicolumn{2}{l|}{Gender} & \multicolumn{2}{l|}{Age} & \multicolumn{2}{l}{Race/Ethnicity}         \\ \hline
Man                & 36.1\% &  25-34       & 19.4\%     & White                                & 61.1\% \\
Woman              & 61.1\% &  35-44  & 55.6 \%     &      Asian  & 27.8\% \\
Non-binary.  & 2.8\%  &      45-54 & 16.7\%     & Black or African American (Non-Hispanic)                            & 5.6\%  \\
&        &     55-64 &    5.6\%  & Hispanic or Latino     & 2.8\%  \\
&        &   65+ &    2.8\%   & Other & 2.8\%  \\\hline
\end{tabular}}
\caption{Breakdown of participant demographics by gender, age, and race/ethnicity - Randomized Trial, mTurk.}
  \vspace{-0.2in}
\label{tab:demographicsMturk2}
\end{table}

\vspace{-5em}